\documentclass[12pt]{article}
\usepackage{booktabs}
\usepackage{tabulary}
\usepackage{multirow}
\usepackage{rotating}
\usepackage{epsfig}
\usepackage{psfrag}
\usepackage{latexsym}
\usepackage{indentfirst}
\usepackage{fancyhdr}
\usepackage{dsfont}
\usepackage{amsmath}
\usepackage{amssymb}
\usepackage{amsfonts}
\usepackage{mathrsfs}
\usepackage{amsthm}
\usepackage{pifont}
\usepackage{dsfont}
\usepackage{multirow}
\usepackage{array}
\usepackage{chngpage}
\usepackage{longtable}
\usepackage{cite}
\usepackage{bbold}
\usepackage{color}
\usepackage{braket}
\usepackage{colordvi}
\usepackage{makecell}
\usepackage{fancybox}
\usepackage[footnotesize]{caption2}
\usepackage{graphicx}
\usepackage[center,footnotesize,hang]{subfigure}
\usepackage{bbm}
\usepackage{url}
\usepackage[colorlinks, linkcolor=red, anchorcolor=black, citecolor=green]{hyperref}
\usepackage{multirow}
\usepackage[integrals]{wasysym}
\usepackage{harpoon}
\usepackage{textcomp}  
\usepackage{enumitem}
\usepackage{adjustbox}
\usepackage{mathrsfs}  
\newcommand{\PreserveBackslash}[1]{\let\temp=\\#1\let\\=\temp}
\newcolumntype{C}[1]{>{\PreserveBackslash\centering}p{#1}}
\newcolumntype{R}[1]{>{\PreserveBackslash\raggedleft}p{#1}}
\newcolumntype{L}[1]{>{\PreserveBackslash\raggedright}p{#1}}
\allowdisplaybreaks  

\newcommand{\bq}{\begin{eqnarray}}
\newcommand{\nq}{\end{eqnarray}}

\newcommand{\cleqn}{\setcounter{equation}{0}}

\begin{document}
\title{
\begin{flushright}
\hfill\mbox{{\small\tt UCI-TR-2022-24}}\\
\hfill\mbox{{\small\tt DESY-22-171}}\\[5mm]
\begin{minipage}{0.2\linewidth}
\normalsize
\end{minipage}
\end{flushright}
{\Large \bf A minimal modular invariant neutrino model \\[2mm]}
\date{}

\author{
Gui-Jun~Ding$^{1}$\footnote{E-mail: {\tt
dinggj@ustc.edu.cn}},  \
Xiang-Gan Liu$^{2}$\footnote{E-mail: {\tt
xianggal@uci.edu}},  \
Chang-Yuan Yao$^{3, 4}$\footnote{E-mail: {\tt
yaocy@nankai.edu.cn}}
\\*[20pt]
\centerline{
\begin{minipage}{\linewidth}
\begin{center}
$^1${\it \small
Department of Modern Physics, University of Science and Technology of China,\\
Hefei, Anhui 230026, China}\\[2mm]
$^2${\it \small Department of Physics and Astronomy, University of California, Irvine, CA 92697-4575, USA}\\[2mm]
$^3${\it \small School of Physics, Nankai University, Tianjin 300071, China}\\[2mm]
$^4${\it \small Deutsches Elektronen-Synchrotron DESY, Notkestr. 85, 22607 Hamburg, Germany}
\end{center}
\end{minipage}}
\\[10mm]}}

\maketitle
\thispagestyle{empty}

\begin{abstract}
We present a neutrino mass model based on modular symmetry with the fewest input parameters to date, which successfully accounts for the 12 lepton masses and mixing parameters through 6 real free parameters including the modulus. The neutrino masses are predicted to be normal ordering, the atmospheric angle $\theta_{23}$ is quite close to maximal value and the Dirac CP phase $\delta_{CP}$ is about $1.34\pi$. We also study the soft supersymmetry breaking terms due to the modulus $F$-term in this minimal model, which are constrained to be the non-holomorphic modular forms. The radiative lepton flavor violation process $\mu\to e\gamma$ is discussed.
\end{abstract}

\section{\label{sec:introduction}Introduction}
The origin of hierarchical fermion masses and flavor mixing parameters is a long-standing puzzle of particle physics, and flavor symmetry has been extensively studied as a guiding principle to understand the flavor puzzle, see Refs.~\cite{Feruglio:2019ybq,King:2017guk,Xing:2020ijf,Almumin:2022rml} for review on this topic. The modular invariance as flavor symmetry was recently proposed to provide a promising framework to address the flavor structure of SM. The Yukawa couplings are constrained to be modular forms of level $N$ which are holomorphic functions of the complex modulus $\tau$, and the flavor symmetry could be uniquely broken down by the vacuum expectation value of $\tau$. The modular flavor symmetry allows to construct predictive flavor models characterized by a small number of Lagrangian parameters, and it is remarkable that all higher-dimensional operators in the superpotential are unambiguously determined in the limit of unbroken supersymmetry (SUSY).

The model construction is based the inhomogeneous finite modular groups $\Gamma_N\equiv\bar{\Gamma}/\bar{\Gamma}(N)$~\cite{Feruglio:2017spp} or homogeneous finite modular groups $\Gamma'_N\equiv\Gamma/\Gamma(N)$~\cite{Liu:2019khw}. For finite modular groups of small order, many lepton and quark mass models have been constructed and discussed, for example $\Gamma_2\cong S_3$~\cite{Kobayashi:2018vbk,Kobayashi:2018wkl}, $\Gamma_3\cong A_4$~\cite{Feruglio:2017spp,Criado:2018thu,Kobayashi:2018vbk, Kobayashi:2018scp,Okada:2018yrn,Kobayashi:2018wkl,Okada:2019uoy,Ding:2019zxk,Kobayashi:2019xvz,Asaka:2019vev,Gui-JunDing:2019wap,Zhang:2019ngf,King:2020qaj,Ding:2020yen,Asaka:2020tmo,Okada:2020brs,Yao:2020qyy,Okada:2021qdf,Nomura:2021yjb,Chen:2021prl,Nomura:2022boj,Gunji:2022xig}, $\Gamma_4\cong S_4$~\cite{Penedo:2018nmg,Novichkov:2018ovf,deMedeirosVarzielas:2019cyj,Kobayashi:2019mna,King:2019vhv,Criado:2019tzk,Wang:2019ovr,Gui-JunDing:2019wap,Wang:2020dbp,Qu:2021jdy}, $\Gamma_5\cong A_5$~\cite{Novichkov:2018nkm,Ding:2019xna,Criado:2019tzk}, $\Gamma_7\cong \mathrm{PSL}(2,\mathbb{Z}_7)$~\cite{Ding:2020msi}, $\Gamma'_3\cong T'$~\cite{Liu:2019khw,Lu:2019vgm,Okada:2022kee}, $\Gamma'_4\cong S'_4$~\cite{Novichkov:2020eep,Liu:2020akv}, $\Gamma'_5\cong A'_5$~\cite{Wang:2020lxk,Yao:2020zml,Behera:2021eut} and $\Gamma'_6\cong S_3\times T'$~\cite{Li:2021buv}.
In the modular invariant models, the Yukawa couplings are integer weight modular forms of the principal congruence subgroup $\Gamma(N)$. Recently, $\Gamma_N$ and $\Gamma'_N$ has been extended to the most general finite modular groups~\cite{Liu:2021gwa}, where the modular forms of $N$ are generalized to be the vector valued modular forms of $\mathrm{SL}(2, \mathbb{Z})$. Moreover, the rational weight modular form in modular flavor symmetry and the metaplectic finite modular groups~$\widetilde{\Gamma}_N\equiv \widetilde{\Gamma}/\widetilde{\Gamma}(N)$ are discussed in Refs.~\cite{Liu:2020msy,Yao:2020zml}. It is known that there are only three independent fixed points $\tau=i$, $e^{2\pi i/3}$, $i\infty$ in the fundamental domain of the modular group~\cite{Gui-JunDing:2019wap}. It has been recognized that the mass hierarchies of the charged leptons can arise from the deviation of the complex modulus from these fixed points~\cite{Okada:2020ukr,Feruglio:2021dte,Novichkov:2021evw,Feruglio:2022kea}. In the top-down approach such as string theory, generally multiple moduli are involved. In view of this, the $\mathrm{SL}(2,\mathbb{Z})$ modular symmetry with single complex modulus has been extended to the $\mathrm{Sp}(2g,\mathbb{Z})$ symplectic modular symmetry and even other modular symmetries in higher dimensional moduli space~\cite{Ding:2020zxw}, where the classical modular forms are replaced by more general automorphic forms. The symplectic modular symmetry would be reduced to the product of several $\mathrm{SL}(2,\mathbb{Z})$ with certain mirror symmetry~\cite{deMedeirosVarzielas:2019cyj,King:2019vhv} when the moduli space is factorized into several independent tori. The predictive power of modular symmetry can be further improved by including the generalized CP symmetry. It is found that the generalized CP symmetry can be consistently combined with symplectic modular symmetries for both single modulus with $g=1$~\cite{Novichkov:2019sqv,Baur:2019kwi,Baur:2019iai} and multiple moduli with $g\geq 2$~\cite{Ding:2021iqp}. The generalized CP symmetry can enforce the coupling constants to be real in certain representation basis, so that the vacuum expectation value (VEV) of $\tau$ is the unique source of flavor symmetry breaking and CP violation. The modular symmetry also can be embed into $\mathrm{SU}(5)$ and $\mathrm{SO}(10)$ Grand Unified Theories~\cite{deAnda:2018ecu,Kobayashi:2019rzp,Du:2020ylx,Zhao:2021jxg,Chen:2021zty,Ding:2021zbg,Ding:2021eva,Charalampous:2021gmf,Ding:2022bzs}. Furthermore, the modular symmetry can naturally appears in top-down constructions~\cite{Baur:2019kwi,Kobayashi:2020hoc,Ohki:2020bpo,Kikuchi:2020frp,Kikuchi:2020nxn,Ishiguro:2020tmo,Nilles:2021glx,Almumin:2021fbk,Kikuchi:2021yog,Baur:2021bly}. However, it is found that the modular symmetry is usually accompanied by traditional flavor symmetry, this scheme is named as eclectic flavor group~\cite{Nilles:2020nnc,Nilles:2020kgo,Baur:2020jwc,Nilles:2020tdp,Nilles:2020gvu}. Some models based on eclectic flavor symmetry have been proposed~\cite{Chen:2021prl,Baur:2022hma}.

The modular flavor symmetry is attractive because the modular invariant mass models usually contain only a few free input parameters, so they have a certain predictive power. Therefore, it is of great significance to find a realistic model with the least free parameters in this framework. As far as we know, the modular invariant models in the literature require at least 7 parameters to explain 12 observables in the lepton sector alone~\cite{Novichkov:2019sqv,Liu:2020akv,Yao:2020qyy,Qu:2021jdy}, and at least 9 parameters to explain 10 observables in the quark sector alone~\cite{Liu:2020akv,Yao:2020qyy,Qu:2021jdy}. In this paper, after searching a large number of possible models, we succeeded in finding a modular invariant lepton model with the fewest parameters so far, which contains only 6 real free parameters but successfully matches the current experimental data. We carefully analyze the best-fit values of this model and perform an exhaustive scan of the parameter space, we find interesting features of the neutrino mixing angles and CP violation phases predicted by this model. On the other hand,  the modular invariant supersymmetry theory constrains not only the flavor structures of quarks and leptons, but also the flavor structures of their superpartners, which leads to specific patterns in soft SUSY breaking terms~\cite{Du:2020ylx,Kobayashi:2021jqu}. These terms will cause lepton flavor violation at low energy. Therefore, we also study the soft SUSY breaking terms in our minimal lepton model and their flavor phenomenological implications such as the branching ratio of rare decay $\mu\to e\gamma$.

This paper is organized as follows. In section~\ref{sec:Framework}, the modular symmetry and the soft SUSY breaking terms in modular invariant supergravity are reviewed. In section~\ref{sec:model}, we present the minimal lepton model which is based on finite modular group $S'_4$, and also analyze the predictions of this model numerically. In section~\ref{sec:LFV}, we show the soft breaking terms in our minimal model, and calculate the branch ratio of lepton flavor violation process $\mu\rightarrow e\gamma$. The finite modular groups $\Gamma'_4\cong S'_4$ and its Clebsch-Gordan coefficients are given in Appendix~\ref{sec:app-S4DC}. In Appendix~\ref{sec:MF}, we give the explicit expressions of the relevant modular forms.

\section{\label{sec:Framework} Modular symmetry and soft terms in supergravity}
We restrict ourselves to the framework of modular invariant supergravity with  single modulus~\cite{Ferrara:1989bc,Feruglio:2017spp}, and consider the moduli-mediated SUSY breaking~\cite{Kaplunovsky:1993rd,Brignole:1993dj,Kobayashi:1994eh,Brignole:2010sax}. The modular symmetry are characterized by the modular group $\Gamma\equiv\mathrm{SL}(2,\mathbb{Z})$ which consists of matrices with integer entries:
\begin{equation}
        \mathrm{SL}(2,\mathbb{Z})=\left\{\begin{pmatrix}
                a &~ b \\ c &~ d
        \end{pmatrix}  \Big|~ ad- bc=1\,,\quad  a,b,c,d \in \mathbb{Z}\,\right\}\,.
\end{equation}
It can be generated by two generators $S$ and $T$
\begin{equation}
        S=\begin{pmatrix}
                0 &~ 1 \\ -1 &~ 0
        \end{pmatrix}\,, \qquad T=\begin{pmatrix}
                1 &~ 1 \\ 0 &~ 1
        \end{pmatrix}\,.
\end{equation}
They obey the following relations
\begin{equation}
\label{eq:SL2Zrelation}
S^4=(ST)^3=1\,, \quad S^2 T=T S^2\,.
\end{equation}
Note that $S^2=-\mathbb{1}_2$, where $\mathbb{1}_2$ denotes the two-dimensional identity matrix. The modular group $\Gamma$ acts on the complex upper half-plane $\mathcal{H}=\{\tau\in\mathbb{C}~|~\text{Im} \tau >0\}$ by linear fractional transformation
\begin{equation}
\gamma \tau \equiv \frac{a\tau+b}{c\tau+d}\,,\qquad \gamma=\begin{pmatrix}
a & b \\c & d
\end{pmatrix}\in \Gamma\,.
\end{equation}
It is easy to find that $\gamma$ and $-\gamma$ give the same action on $\tau$ and thus the faithful action of the linear fractional transformation is given by the projective special linear group $\mathrm{PSL}(2,\mathbb{Z})$ which is the quotient group $\mathrm{PSL}(2,\mathbb{Z})\cong \mathrm{SL}(2,\mathbb{Z})/\{\pm\mathbb{1}_2\}$.

The action of modular group on the matter fields $\Phi_i$ is assumed as follows
\begin{equation}
\Phi_i \stackrel{\gamma}{\longrightarrow} (c\tau+d)^{-k_i} \rho(\gamma) \Phi_i\,,
\end{equation}
where $-k_i$ is called the modular weight of matter field $\Phi_i$, and $\rho(\gamma)$ is the unitary irreducible representation of $\mathrm{SL}(2,\mathbb{Z})$ with finite image~\cite{Liu:2021gwa}. In general, the representation can be taken for cases where its kernels are principal congruence subgroups $\Gamma(N)$, and the $\rho(\gamma)$ is often referred to as the representation of finite modular groups $\Gamma'_N\equiv\Gamma/\Gamma(N) $.

The full $\mathcal{N}=1$ supergravity Lagrangian is specified in terms of  two functions: the gauge kinetic function $f_a$ and the real gauge-invariant K\"ahler function $\mathcal{G}(\tau,\Phi_i;\bar{\tau},\bar{\Phi}_i)$~\footnote{We use the standard supergravity mass units, namely the reduced Planck mass $M_p\equiv M_{Planck}/\sqrt{8\pi}=1$.}. $f_a$ determines the kinetic terms for the fields in the vector multiplets and in particular the gauge coupling constant, $\text{Re}f_a=1/g^2_a$, where the subscript $a$ is associated with the different gauge groups of the theory. The K\"ahler function is the combination~\cite{Kaplunovsky:1993rd,Brignole:2010sax}:
\begin{equation}
\mathcal{G}(\tau,\Phi_i;\bar{\tau},\bar{\Phi}_i)=\mathcal{K}(\tau,\Phi_i;\bar{\tau},\bar{\Phi}_i)+\log \mathcal{W}(\tau,\Phi_i)+ \log \mathcal{\overline{W}}(\bar{\tau},\bar{\Phi}_i)\,,
\end{equation}
where the K\"ahler  potential and superpotential have the form
\begin{align}
\mathcal{K}&=\mathcal{K}_\tau + \mathcal{K}_{matter}
+ \dots \,,\\
\mathcal{W}&=Y_{ijk}(\tau)\Phi^i\Phi^j\Phi^k + \mu_{ij}(\tau) \Phi^i \Phi^j + \dots \,.
\label{eq:sup}
\end{align}
In particular, we take the minimal form of K\"ahler  potential~\cite{Feruglio:2017spp}
\begin{equation}
\label{eq:Kah-sup}
\mathcal{K}_\tau=-\log (-i(\tau-\bar{\tau}))\,,\quad\mathcal{K}_{matter}= \tilde{\mathcal{K}}_{i\bar{i}}|\Phi^i|^2=(-i(\tau-\bar{\tau}))^{-k_i}|\Phi^i|^2\,,
\end{equation}
where $\tilde{\mathcal{K}}_{i\bar{i}}$ is the K\"ahler metric. As you can see, the transformation of the K\"ahler  potential induced by the modular transformation of fields is exactly a K\"ahler transformation:
\begin{equation}
\mathcal{K} \stackrel{\gamma}{\longrightarrow}  \mathcal{K} + \log (c\tau+d) +\log (c\bar{\tau}+d)\,.
\end{equation}
Hence, the modular invariance of K\"ahler function $\mathcal{G}$ requires that the superpotential must be transformed complementally:
\begin{equation}
\mathcal{W} \stackrel{\gamma}{\longrightarrow} (c\tau+d)^{-1} \mathcal{W}\,.
\end{equation}
In other words, the superpotential behaves like a chiral superfield with modular weight $-1$. Modular invariance requires that the Yukawa couplings $Y_{ijk}(\tau)$ in Eq.~\eqref{eq:sup} should be modular forms of weight $k_Y$, specifically,
\begin{equation}
Y_{ijk}(\tau) \stackrel{\gamma}{\longrightarrow} Y_{ijk}(\gamma\tau) = (c\tau+d)^{k_Y} \rho(\gamma)_{(ijk)(lmn)} Y_{lmn}(\tau)
\end{equation}
with $k_Y=k_i+k_j+k_k-1$ and product $\rho \times \rho_i \times\rho_j \times\rho_k$ contains an invariant singlet.

The scalar component of modulus in the hidden sector, $\tau$, may obtain large VEV that induce SUSY breaking via non-vanishing VEV of its auxiliary field $F^\tau$.
The gravition becomes massive and its mass is given by~\cite{Kaplunovsky:1993rd,Brignole:2010sax}
\begin{equation}
m_{3/2}=e^{\mathcal{G}/2}M_p\,.
\end{equation}
On the other hand, after taking the so-called flat limit where $M_p\to \infty$ but $m_{3/2}$ is kept fixed, all that is left in the observable sector is a effective global SUSY Lagrangian plus a set of soft SUSY-breaking terms.
The effective superpotential is given by ~\cite{Kaplunovsky:1993rd,Brignole:2010sax}
\begin{equation}
\mathcal{W}^{(eff)}(\Phi_i)= \hat{Y}_{ijk}(\tau)\hat{\Phi}^i\hat{\Phi}^j\hat{\Phi}^k + \hat{\mu}_{ij}(\tau) \hat{\Phi}^i \hat{\Phi}^j
\end{equation}
in the canonically normalized basis,
with normalized Yukawa couplings and normalized masses
\begin{equation}
\hat{Y}_{ijk}= Y_{ijk} e^{\mathcal{K}_{\tau}/2}\left(\tilde{\mathcal{K}}_{i\bar{i}} \tilde{\mathcal{K}}_{j\bar{j}} \tilde{\mathcal{K}}_{k\bar{k}}\right)^{-1/2} \,,\quad
\hat{\mu}_{ij}=\mu_{ij}  e^{\mathcal{K}_{\tau}/2}\left(\tilde{\mathcal{K}}_{i\bar{i}} \tilde{\mathcal{K}}_{j\bar{j}} \right)^{-1/2} \,.
\end{equation}
The effective soft SUSY-breaking Lagrangian in the canonically normalized basis is given by~\cite{Kaplunovsky:1993rd,Brignole:2010sax}
\begin{equation}
\mathcal{L}_{soft}=\frac{1}{2} \left(M_a \hat{\lambda}^a \hat{\lambda}^a +\mathrm{h.c.}\right) - \tilde{m}^2_i \bar{\hat{\Phi}}^i \hat{\Phi}^i -\left(A_{ijk}\hat{Y}_{ijk} \hat{\Phi}^i  \hat{\Phi}^j  \hat{\Phi}^k + B\hat{\mu} \hat{H}_u \hat{H}_d + \mathrm{h.c.} \right)
\end{equation}
with
\begin{equation}
\label{eq:general_soft}
\begin{aligned}
\tilde{m}_{i}^{2} &=m_{3/2}^{2}-|F^{\tau}|^2 \partial_{\bar{\tau}} \partial_{\tau} \log \tilde{\mathcal{K}}_{i\bar{i}}\,, \\
M_{a} &=\frac{1}{2}\left(\mathrm{Re}f_{a}\right)^{-1} F^{\tau} \partial_{\tau} f_{a}\,, \\
A_{ijk} &=F^{\tau}\left[\partial_{\tau} \mathcal{K}_{\tau}+\partial_{\tau} \log Y_{ijk}-\partial_{\tau} \log \left(\tilde{\mathcal{K}}_{i\bar{i}} \tilde{\mathcal{K}}_{j\bar{j}} \tilde{\mathcal{K}}_{k\bar{k}}\right)\right]\,, \\
B &=F^{\tau}\left[\partial_{\tau} \mathcal{K}_{\tau}+\partial_{\tau} \log \mu-\partial_{\tau} \log \left(\tilde{\mathcal{K}}_{H_u} \tilde{\mathcal{K}}_{H_d}\right)\right]-m_{3/2}\,,
\end{aligned}
\end{equation}
where $\tilde{\mathcal{K}}_{H_{u,d}}=(-i(\tau-\bar{\tau}))^{-k_{H_{u,d}}}$,  $\hat{\Phi}^i$ and $\hat{\lambda}^a $ are the scalar and gaugino canonically normalized fields respectively
\begin{equation}
\hat{\Phi}^i= \tilde{\mathcal{K}}_{i\bar{i}}^{1/2} \Phi^i \,,\quad \hat{\lambda}^a = (\mathrm{Re}f_a)^{1/2} \lambda^a \,.
\end{equation}

\section{\label{sec:model} A minimal neutrino mass model based on $S'_4$ modular symmetry}

In this section, we shall present a model for neutrino masses and mixing based on the $S'_4$ modular symmetry, and it depends on only six real parameters including the modulus $\tau$. It is the phenomenologically viable lepton mass model with the smallest number of free parameters as far as we know. The generalized CP (gCP) symmetry has been included in this model in order to increase the predictive power. It is known the complex modulus $\tau$ transforms as $\tau\rightarrow -\tau^{*}$ under the action of gCP. In the symmetric basis where both modular generator $S$ and $T$ are represented by symmetric and unitary matrices, gCP reduces to canonical CP transformation~\cite{Novichkov:2019sqv,Ding:2021iqp}. As a consequence, the gCP symmetry would constrain all couplings constants to be real in the representation basis with real Clebsch-Gordan coefficients. As shown in the Appendix~\ref{sec:app-S4DC}, we indeed works in symmetric basis of $S'_4$ and the all the Clebsch-Gordan coefficients are real.

In this model, the neutrino masses are described by type-I seesaw mechanism. We introduce three right-handed neutrinos $N^c=(N^c_1, N^c_2, N^c_3)^T$ and assume that they transform according to the triplet $\mathbf{3}$ of $S'_4$. In charged lepton sector, the first two generations of the right-handed charged leptons $E^c_{D}=(E^c_1, E^c_2)^{T}$ are assigned to the doublet representation $\mathbf{\hat{2}}$, and the third generation of the right-handed charged leptons $E^c_3$ is assigned to be $S'_4$ singlet. The left-handed charged lepton $L=(L_1,L_2,L_3)^T$ transforms as a triplet $\mathbf{3}$. The representation and weight assignments of the fields are summarized as follows:
\begin{align}
        \nonumber&\rho_{E^c}=\mathbf{\hat{2}}\oplus\mathbf{\hat{1}'},~ \quad \rho_{L} = \mathbf{3},\quad ~\rho_{N^c} = \mathbf{3},~\quad \rho_{H_u}=\rho_{H_d}=\mathbf{1}\,,\\
        &k_{E_{1,2,3}^c}=11/2,11/2,11/2\,,~\quad k_{N^c}=3/2,~ \quad k_{L}=-3/2\,,~\quad k_{H_u}=k_{H_d}=0\,.
\end{align}
The superpotential of the lepton sector includes:
\begin{align}
\nonumber \mathcal{W}_e &=  \alpha \left( E_D^c L Y^{(3)}_{\mathbf{\hat{3}'}}\right)_{\mathbf{1}} H_d + \beta \left( E_D^c L Y^{(3)}_{\mathbf{\hat{3}}}\right)_{\mathbf{1}} H_d + \gamma \left(E_3^c L Y^{(3)}_{\mathbf{\hat{3}}}\right)_{\mathbf{1}} H_d\,, \\
\mathcal{W}_\nu &=  g_1 \left(N^c L\right)_{\mathbf{1}} H_u + \Lambda \left( (N^c N^c)_{\mathbf{2},s} Y^{(2)}_{\mathbf{2}}\right)_{\mathbf{1}}\,.
\end{align}
Then, the charged lepton and neutrino mass matrices can be read off by using the Clebsch-Gordon coefficients of $S'_4$ shown in Appendix~\ref{sec:app-S4DC}:
\begin{align}
\label{eq:S4DClept1}
\nonumber&M_e=  \begin{pmatrix}
2\alpha Y^{(3)}_{\mathbf{\hat{3}'},1} ~&~ -\alpha Y^{(3)}_{\mathbf{\hat{3}'},3}+\sqrt{3}\beta Y^{(3)}_{\mathbf{\hat{3}},2}   ~&~  -\alpha Y^{(3)}_{\mathbf{\hat{3}'},2}+\sqrt{3}\beta Y^{(3)}_{\mathbf{\hat{3}},3}  \\
-2 \beta Y^{(3)}_{\mathbf{\hat{3}},1} ~&~ \sqrt{3}\alpha Y^{(3)}_{\mathbf{\hat{3}'},2}+\beta Y^{(3)}_{\mathbf{\hat{3}},3}  ~&~ \sqrt{3}\alpha Y^{(3)}_{\mathbf{\hat{3}'},3}+\beta Y^{(3)}_{\mathbf{\hat{3}},2} \\
 \gamma Y^{(3)}_{\mathbf{\hat{3}},1} ~&~ \gamma Y^{(3)}_{\mathbf{\hat{3}},3} ~&~ \gamma Y^{(3)}_{\mathbf{\hat{3}},2}
\end{pmatrix}v_d  \,, \\
&M_D=g\begin{pmatrix}
        1  & 0 & 0 \\
    0  & 0 & 1 \\
        0  & 1 & 0
\end{pmatrix}v_u\,,\quad M_N= \Lambda \begin{pmatrix}
2 Y^{(2)}_{\mathbf{2},1} & 0 & 0  \\
0 & \sqrt{3} Y^{(2)}_{\mathbf{2},2} & -Y^{(2)}_{\mathbf{2},1} \\
0 & -Y^{(2)}_{\mathbf{2},1} & \sqrt{3} Y^{(2)}_{\mathbf{2},2}
        \end{pmatrix}\,.
\end{align}
The light neutrino mass matrix $M_{\nu}$ is given by the seesaw formula
\begin{equation}
M_\nu= -M_D^T M_N^{-1} M_D = \frac{g^2v_u^2}{\Lambda}\begin{pmatrix}
-\frac{1}{2 Y^{(2)}_{\mathbf{2},1}} ~&~ 0 ~&~ 0  \\
0 ~&~ \frac{\sqrt{3} \; Y^{(2)}_{\mathbf{2},2}}{Y^{(2)2}_{\mathbf{2},1}-3Y^{(2)2}_{\mathbf{2},2}} ~&~ \frac{Y^{(2)}_{\mathbf{2},1}}{Y^{(2)2}_{\mathbf{2},1}-3Y^{(2)2}_{\mathbf{2},2}} \\
0 ~&~ \frac{Y^{(2)}_{\mathbf{2},1}}{Y^{(2)2}_{\mathbf{2},1}-3Y^{(2)2}_{\mathbf{2},2}} ~&~ \frac{\sqrt{3}\; Y^{(2)}_{\mathbf{2},2}}{Y^{(2)2}_{\mathbf{2},1}-3Y^{(2)2}_{\mathbf{2},2}}
\end{pmatrix}\,.
\end{equation}
It is remarkable that the light neutrino mass matrix $M_\nu$ is a block diagonal matrix, and consequently we can easily read off the light neutrino masses as follows:
\begin{equation}
\label{eq:NeutrinoMass}
m_1=\frac{1}{|2Y^{(2)}_{\mathbf{2},1}|} \frac{g^2v_u^2}{\Lambda}\,,\quad m_2=\frac{1}{|Y^{(2)}_{\mathbf{2},1}-\sqrt{3}Y^{(2)}_{\mathbf{2},2}|}\frac{g^2v_u^2}{\Lambda}\,,\quad
m_3=\frac{1}{|Y^{(2)}_{\mathbf{2},1}+\sqrt{3}Y^{(2)}_{\mathbf{2},2}|}\frac{g^2v_u^2}{\Lambda}\,.
\end{equation}
In the modular invariant models, the determinant of the lepton mass matrices are some one-dimensional vector-valued modular forms of $\mathrm{SL}(2,\mathbb{Z})$~\cite{Liu:2021gwa}. In this minimal model, we have
\begin{align}
\label{eq:massdet}
&\det[M_e(\tau)]=-96\sqrt{6}v_d^3\gamma(\beta^2-3\alpha^2)\eta^{18}(\tau)\,,
\end{align}
where $\eta(\tau)=e^{\pi i \tau/12} \prod_{n=1}^{\infty} (1-e^{2\pi i n\tau})$ is the well-known Dedekind eta function. We see that the small electron mass can be naturally reproduced for $\beta\approx\pm\sqrt{3}\alpha$.

\begin{figure}[t!]
\centering
\includegraphics[width=0.65\textwidth]{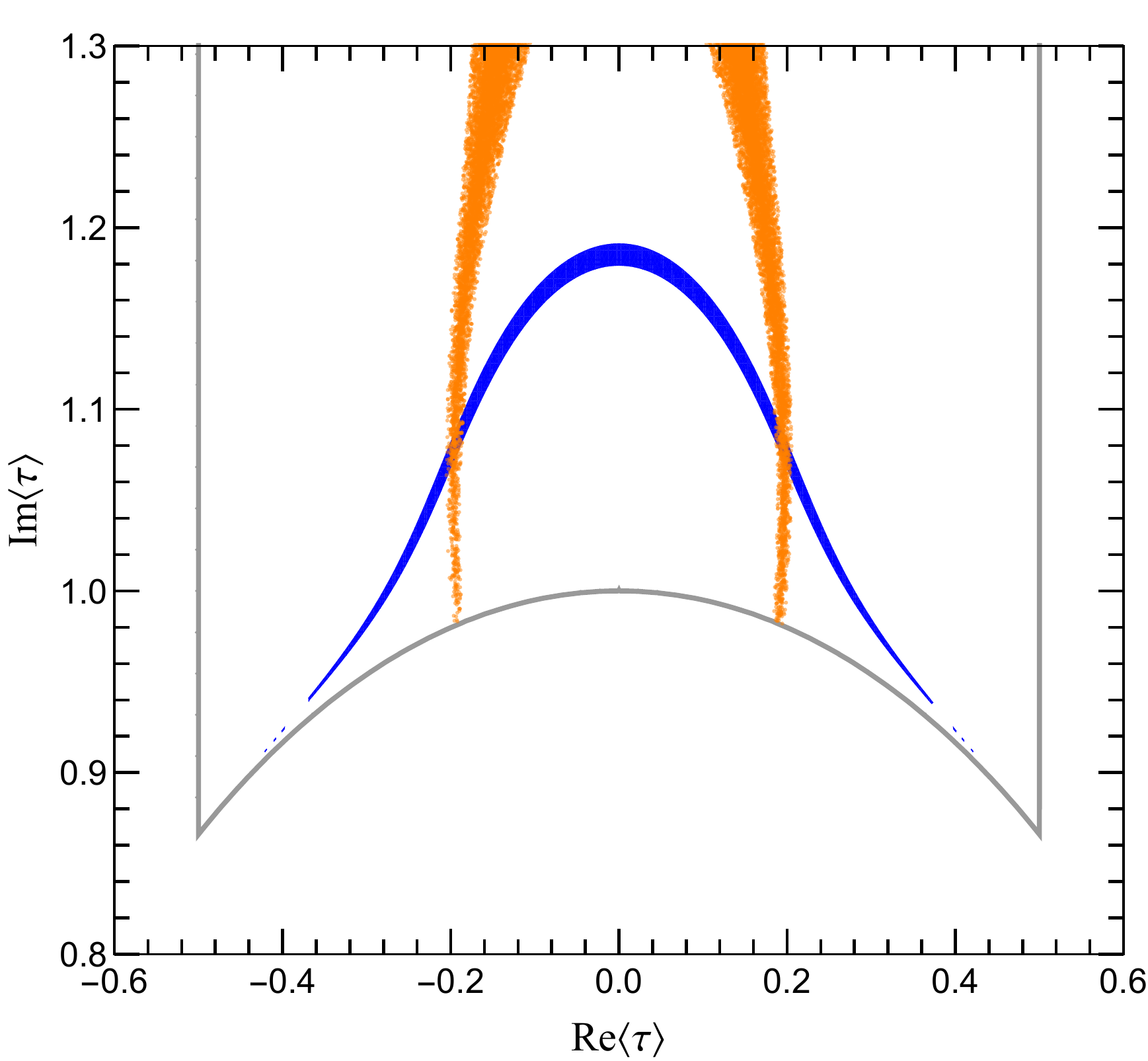}
\caption{The region of modulus $\tau$ compatible with experimental data, where the gray line is the boundary of the fundamental domain. The blue region represents the feasible range of $\langle\tau\rangle$ compatible with the data $\Delta m^2_{21}/\Delta m^2_{31}$ of the neutrino mass squared difference~\cite{Esteban:2020cvm}. The orange area denotes  the viable region of $\langle\tau\rangle$ limited only by the measured values of the charged lepton mass ratios and the reactor mixing angle $\theta_{13}$~\cite{Antusch:2013jca,Esteban:2020cvm}. }
\label{fig:tauRegion}
\end{figure}

As explained in the beginning of this section, the gCP symmetry constrains all couplings to be real in our working basis. Thus,
all lepton flavor observable only depend on four coupling constants $\alpha$, $\beta$, $\gamma$, $g^2/\Lambda$ plus the complex modulus $\langle \tau\rangle$ in our model. A notable feature of this model is that the light neutrino mass matrix as well as the neutrino mass ratios are completely determined by the modulus $\tau$ up to the overall scale $g^2v^2_u/\Lambda$. After numerical fitting, we find the experimental data can be accommodated only if the neutrino masses are normal ordering, and the best-fit values of the input parameters that agree well with the experimental data are given by~\cite{Esteban:2020cvm}:
\begin{equation}
\begin{gathered}
\label{eq:best-fit values}
\langle \tau \rangle =-0.193773+1.08321i\,,\quad \beta/\alpha= 1.73048\,,\quad \gamma/\alpha=0.27031\,,\\
\alpha v_d = 244.621~ \text{MeV}\,,\quad g^2v^2_u/\Lambda= 29.0744~\text{meV}\,,
\end{gathered}
\end{equation}
where $\alpha v_d$ and $g^2v^2_u/\Lambda$ are fixed by the measured values of the electron mass and the solar mass squared splitting $\Delta m^2_{21}$ respectively~\cite{Esteban:2020cvm}. It is worth noting that all dimensionless input parameters happen to be $\mathcal{O}(1)$, and $\beta/\alpha$ is close to $\sqrt{3}$. Notice that the electron mass is exactly vanishing when $\beta/\alpha=\sqrt{3}$, see Eq.~\eqref{eq:massdet}.
At the above best fit point, the charged lepton mass ratios, the lepton mixing angles, CP violating phases and the neutrino masses are determined to be:
\begin{equation}
\begin{gathered}
\sin^2\theta_{12}=0.328920\,,\quad \sin^2\theta_{13}=0.0218499\,,\quad \sin^2\theta_{23}=0.506956\,,\quad \delta_{CP}=1.34256\pi\,,\\
\alpha_{21}= 1.32868\pi \,,~~\alpha_{31}= 0.544383\pi \,,~~ m_e/m_\mu=0.00472633 ,~~m_\mu/m_\tau=0.0587566\,,  \\
m_1=14.4007~\text{meV}\,,\quad m_2 =16.7803~\text{meV}\,,\quad m_3 =51.7755~\text{meV}\,,\\
m_\beta=16.8907~\text{meV} \,,\quad m_{\beta\beta}=9.25333~\text{meV}\,,
\end{gathered}
\end{equation}
where $m_\beta$ is the effective neutrino masses probed by direct kinematic search in tritium beta decay and $m_{\beta\beta}$ is the effective mass in neutrinoless double beta decay. We see that the neutrino mass sum is predicted to be $m_1+m_2+m_3=82.9565$ meV which is compatible with the upper limit of Planck $\sum_i m_i < 120$ meV~\cite{Aghanim:2018eyx}. We would like to emphasize that inverted neutrino mass ordering is disfavored in our model. The predicted neutrino mixing angles and CP violation phase $\delta_{CP}$ are within the $3\sigma$ intervals of the latest global fit NuFIT v5.1 without SK atmospheric data~\cite{Esteban:2020cvm}, the charged lepton mass ratios are compatible with their renormalization group (RG) running values at the GUT
scale $2\times 10^{16}$~GeV, where $M_{\text{SUSY}}=1~\mathrm{TeV}\,,\tan\beta=5$ is taken as a benchmark~\cite{Antusch:2013jca}.

\begin{figure}[t!]
\centering
\includegraphics[width=1.0\textwidth]{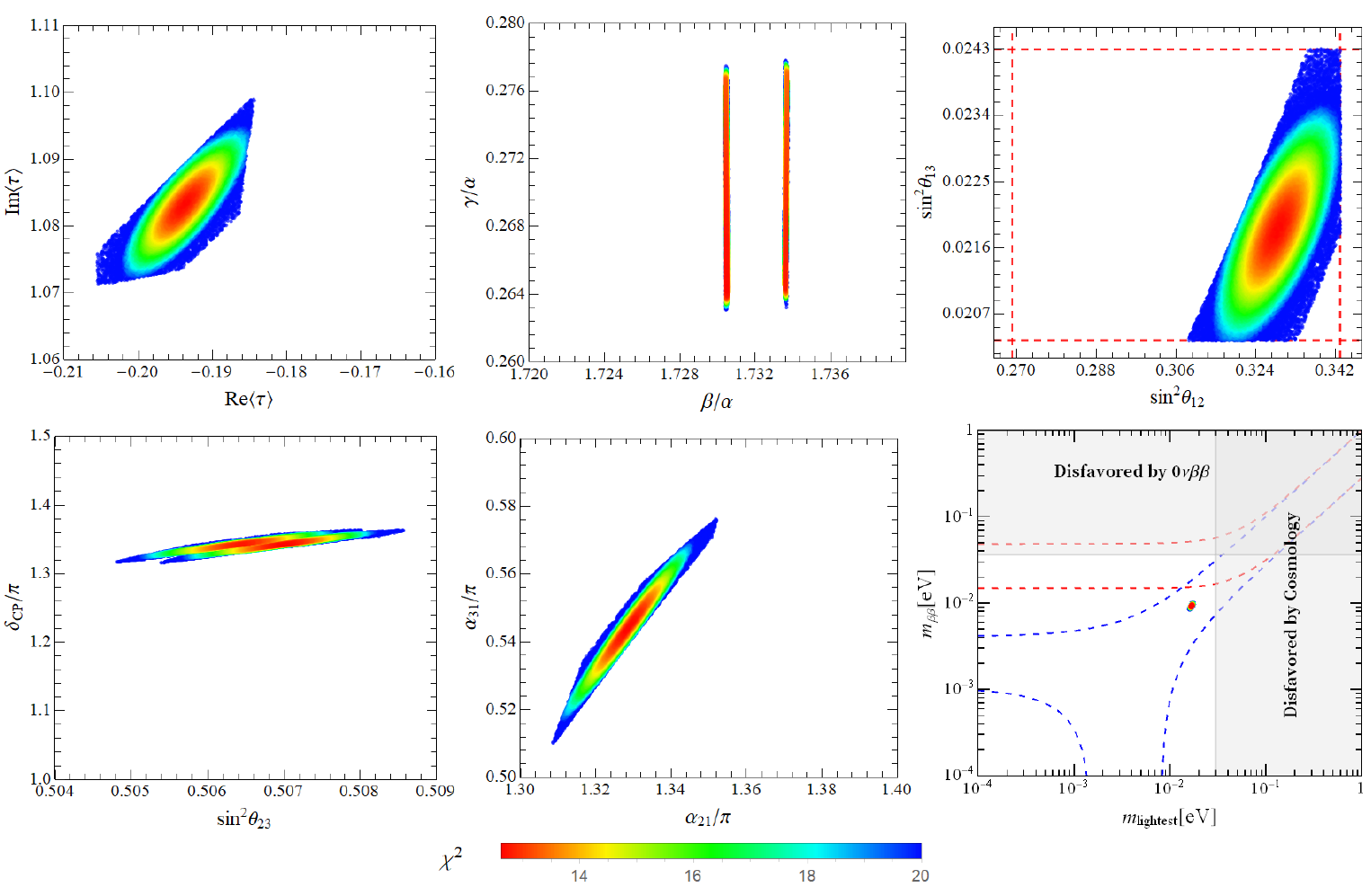}
\caption{The predicted correlations among the input free parameters, neutrino mixing angles and CP violation phases in the minimal model. The plots only display the points which can reproduce the charged lepton masses, $\Delta m^2_{21}$, $\Delta m^2_{31}$ and all the three lepton mixing angles at $3\sigma$ level~\cite{Esteban:2020cvm}. In the top-right panel, the red dashed line are the $3\sigma$ bounds of the mixing angles. In the bottom right panel for $m_{\beta\beta}$, the blue (red) dashed lines represent the most general allowed regions for normal ordering (inverted ordering) neutrino mass spectrum, where the neutrino oscillation parameter are
varied within their $3\sigma$ ranges. Moreover, the vertical grey exclusion band stands for the most radical upper bound $\sum_i m_i<0.12~\text{eV}$ form Planck~\cite{Aghanim:2018eyx}. The horizontal grey band represents the present upper limit $m_{\beta\beta}\leq (36-156)$ meV from KamLAND-Zen~\cite{KamLAND-Zen:2022tow}.}
\label{fig:correlations}
\end{figure}

From Eq.~\eqref{eq:NeutrinoMass}, we see that the light neutrino masses only depends on the VEV of the modulus $\tau$ and the overall mass scale $g^2v^2_u/\Lambda$. Hence we can use the measure value of the ratio $\Delta m^2_{21}/\Delta m^2_{31}$ to constrain the range of $\langle\tau\rangle$, where $\Delta m^2_{21}\equiv m^2_2-m^2_1$ and $\Delta m^2_{31}\equiv m^2_3-m^2_1$ are the solar and atmospheric neutrino mass squared differences respectively. The corresponding result is shown in the blue region of figure~\ref{fig:tauRegion}. Furthermore, we use the precisely measured values of the reactor angle $\theta_{13}$ and the charged lepton mass ratios $m_e/m_{\mu}$, $m_{\mu}/m_{\tau}$ to limit the phenomenologically allowed region of $\langle\tau\rangle$, and the both parameters $\beta/\alpha$ and $\gamma/\alpha$ are allowed to freely vary. The corresponding the result is displayed by the orange area in figure~\ref{fig:tauRegion}. Therefore the modulus should lie in two small regions around $-0.19+1.08i$ and $0.19+1.08i$ in order to accommodate the current data. Moreover, we also have comprehensively explored the parameter space of this minimal model. Requiring the three charged lepton masses $m_{e, \mu, \tau}$, the three lepton mixing angles $\theta_{12}$, $\theta_{13}$, $\theta_{23}$ and the neutrino squared
mass splittings $\Delta m^2_{21}$ and $\Delta m^2_{31}$ to lie in the experimentally allowed $3\sigma$ regions~\cite{Esteban:2020cvm}, we get the
correlations between the free parameters and observable quantities, which are shown in figure~\ref{fig:correlations}. If $\langle\tau\rangle$ is changed to $-\langle\tau\rangle^*$ and the values of coupling constants are kept intact, the sign of the CP violation phases $\delta_{CP}$, $\alpha_{21}$, $\alpha_{31}$ would be reversed while predictions for lepton masses and mixng angles remain the same. As a consequence, we only plot the region of  $\text{Re}\langle\tau\rangle<0$ for simplicity. It can be seen that the overlapping region in figure~\ref{fig:tauRegion} almost coincides with the $\langle\tau\rangle$ region shown in figure~\ref{fig:correlations}. It is remarkable that the phenomenologically viable parameter space is actually very small.

In particular, we notice that the neutrino mixing angle $\sin^2\theta_{23}$ is limited in the range of $0.504$ and $0.510$ which is in the second octant, the Dirac CP violation phase $\delta_{CP}$ lie in a very small interval $[1.316\pi, 1.364\pi]$. These predictions for $\theta_{23}$ and $\delta_{CP}$ could be tested in forthcoming long baseline neutrino experiments DUNE~\cite{DUNE:2015lol} and T2HK~\cite{Hyper-KamiokandeProto-:2015xww}. In addition, the neutrino mixing angles $\sin^2\theta_{12}$ and $\sin^2\theta_{13}$ also show a certain correlation, this feature is expected to be tested at JUNO~\cite{JUNO:2015sjr} which can measure the solar angle $\theta_{12}$ with sub-percent precision. Moreover, the Majorana CP violation phases are also found to lie in quite narrow regions $\alpha_{21}\in [1.309\pi, 1.352\pi]$ and $\alpha_{31}\in[0.510\pi, 0.576\pi]$. Consequently we have definite prediction for the effective Majorana mass $m_{\beta\beta}$ in the interval $[8.543~\text{meV}, 10.010~\text{meV}]$ which is within the reach of future ton-scale neutrinoless double beta decay experiments.

\section{\label{sec:LFV}Lepton flavor violation in the minimal modular model}

The SUSY flavor phenomena of lepton flavor violations (LFV) for lepton sector have been discussed in the traditional flavor symmetry models~\cite{Feruglio:2009hu,Ishimori:2010su,Dimou:2015cmw} and modular flavor models~\cite{Kobayashi:2021jqu,Tanimoto:2021ehw}. In this section we will discuss the SUSY flavor phenomena in our minimal modular lepton model.

From the general results of the K\"ahler potential Eq.~\eqref{eq:Kah-sup} and the soft terms Eq.~\eqref{eq:general_soft}, we can obtain the expressions of the soft mass and $A$-term coefficient as follow~\footnote{Interestingly, the soft mass $\tilde{m}_i^2$ is modular invariant because gravitino mass $m_{3/2}$ is modular invariant and $F^\tau$ has modular weight $-2$. The $A$-term $h_{ijk}\equiv Y_{ijk}A_{ijk}$ transform in the same way as $Y_{ijk}$ and they are the non-holomorphic modular forms of weight $k_{Y}$~\cite{Kikuchi:2022pkd}.}:
\begin{align}
\label{eq:soft mass A}
\tilde{m}_i^2 &= m^2_{3/2} -k_i \frac{|F^\tau|^2}{\left(2\mathrm{Im}\tau\right)^2}\,,\\
A_{ijk}&=i(k_i+k_j+k_k-1)\frac{F^\tau}{2\mathrm{Im}\tau} - \frac{F^\tau}{Y_{ijk}}\frac{d Y_{ijk}}{d \tau}\,.
\end{align}
In order to estimate the magnitude of the flavor changing neutral current (FCNC). We take the so-called mass insertion (MI) approximation, and
move to the super-CKM (SCKM) basis, i.e., the basis where through a rotation of the whole superfield (fermion + sfermion), we obtain diagonal Yukawa couplings for the corresponding fermion fields. The mass insertion parameters $\left(\delta^{\ell}_{LL}\right)_{ij},\left(\delta^{\ell}_{LR}\right)_{ij},\left(\delta^{\ell}_{RL}\right)_{ij}$ and $\left(\delta^{\ell}_{RR}\right)_{ij}$ are defined by
\begin{equation}
\begin{pmatrix}
\tilde{m}_{eL}^2 &      \tilde{m}_{eLR}^2 \\
\tilde{m}_{eRL}^2 &     \tilde{m}_{eRR}^2
\end{pmatrix} = m_{\tilde{\ell}}^{2} \begin{pmatrix}
\delta^{\ell}_{LL} &    \delta^{\ell}_{LR} \\
\delta^{\ell}_{RL} &    \delta^{\ell}_{RR}
\end{pmatrix}  + \text{diag}\left( m_{\tilde{\ell}}^{2} \right)\,,
\end{equation}
where $ m_{\tilde{\ell}}$ refers to the average slepton mass, and
\begin{align}
\label{eq:SleptonMass}
\nonumber&\tilde{m}_{eL}^2=\text{diag}\left(  m^2_{3/2}+\frac{3}{2}\frac{|F^\tau|^2}{\left(2\mathrm{Im}\tau\right)^2},m^2_{3/2}+\frac{3}{2}\frac{|F^\tau|^2}{\left(2\mathrm{Im}\tau\right)^2},m^2_{3/2}+\frac{3}{2}\frac{|F^\tau|^2}{\left(2\mathrm{Im}\tau\right)^2}\right)\,, \\
\nonumber& \tilde{m}_{eR}^2=\text{diag}\left(m^2_{3/2}-\frac{11}{2}\frac{|F^\tau|^2}{\left(2\mathrm{Im}\tau\right)^2},m^2_{3/2}-\frac{11}{2}\frac{|F^\tau|^2}{\left(2\mathrm{Im}\tau\right)^2},m^2_{3/2}-\frac{11}{2}\frac{|F^\tau|^2}{\left(2\mathrm{Im}\tau\right)^2}\right)\,, \\
\nonumber&\tilde{m}_{eRL}^2 =v_d A_{ijk}Y_{ijk}= -  F^\tau\left(\frac{d}{d\tau}-i\frac{3}{2\mathrm{Im}\tau}  \right) M_e \,,\\
&\tilde{m}_{eLR}^2=\tilde{m}_{eRL}^{2\,\dagger}\,.
\end{align}
As we can see, the soft mass $\tilde{m}_i$ is flavor blind due to the common weights for three generations, therefore, only the $A$-term contributes to the LFV.

\begin{figure}[ht!]
\centering
\includegraphics[width=0.65\textwidth]{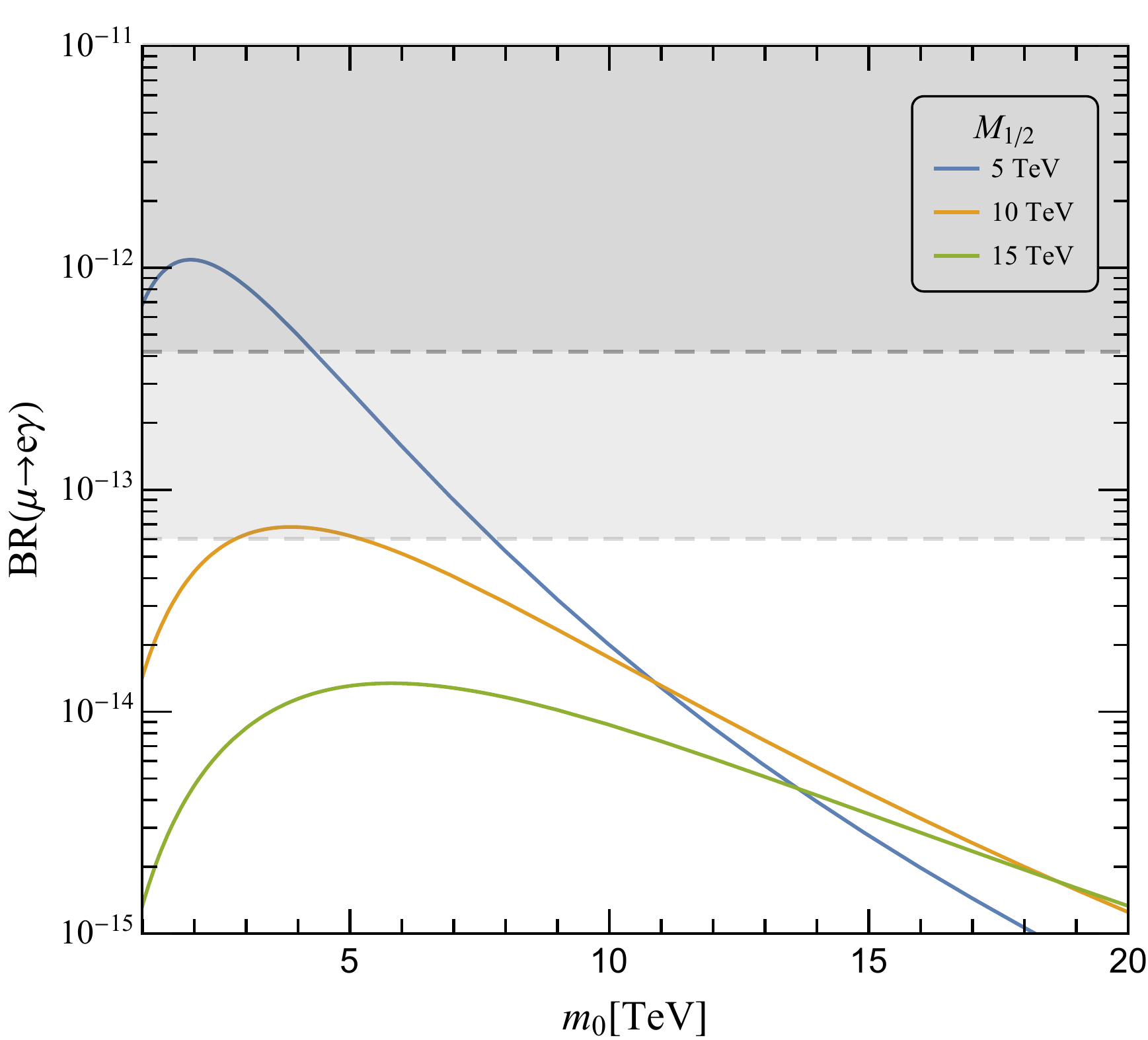}
\caption{The prediction for $\mathrm{BR}(\mu\to e\gamma)$ versus $m_0$ with $F^\tau=m_0/4$ in the minimal model for $M_{1/2}=5, 10, 15~\mathrm{TeV}$ respectively. The dark grey region is excluded by the current experimental bound $\mathrm{BR}(\mu\to e\gamma)< 4.2\times 10^{-13}$~\cite{MEG:2016leq}. The light grey dashed line denotes the future expected bound~\cite{MEGII:2018kmf}.}
\label{fig:LFV}
\end{figure}

Note that our model are defined at high energy scales $Q_0$ (for example, GUT scale), so in order to analyze the phenomenology of these quantities at low energy scale $Q$ (for example 1 TeV), we need to consider the effects of their RG running. We take $\tan\beta=5$, then the largest contributions to the elements of the $A$-term arise from those of gauge couplings, we can estimate the running effects by~\cite{Martin:1993zk,Martin:1997ns}
\begin{equation}
A_{ijk}(Q) \simeq e^{-\frac{1}{16\pi^2} \int_{Q_0}^{Q} dt (\frac{9}{5}g_1^2 +3 g_2^2)} A_{ijk}(Q_0)\approx 1.4  A_{ijk}(Q_0)\,,
\end{equation}
where $g_{1,2}$ are the $\mathrm{SU}(2)_L\times \mathrm{U}(1)_Y$ gauge couplings and $t=\log Q/Q_0$, we take $Q_0=10^{16}\,\mathrm{TeV}, Q=1\,\mathrm{TeV}$.
We denote the common scale of soft mass for all scalar particles by $m_0$, and the common scale of gauginos masses by $M_{1/2}$, these two parameters are not fixed by modular flavor symmetry. At $Q_0$, we take the bino mass $M_1$ and wino mass $M_2$ as
\begin{equation}
M_1(Q_0)=M_2(Q_0)=M_{1/2}\,.
\end{equation}
The RG effects lead to the following gauginos masses at low energy scale $Q$~~\cite{Martin:1993zk,Martin:1997ns}:
\begin{equation}
M_1(Q)\simeq \frac{\alpha_1(Q)}{\alpha_1(Q_0)} M_1(Q_0)\,,\quad M_2(Q)\simeq \frac{\alpha_2(Q)}{\alpha_2(Q_0)} M_1(Q_0)\,,
\end{equation}
where $\alpha_i=g_i^2/4\pi$ and $\alpha_1(Q_0)=\alpha_2(Q_0)\simeq 1/25$ at GUT scale $Q_0=10^{16}\,\mathrm{GeV}$. At low energy scale $Q=1\,\mathrm{TeV}$ we have
\begin{equation}
M_1=0.49 M_{1/2}\,,\quad M_2=0.86 M_{1/2}\,.
\end{equation}
The amplitude of rare decay $\ell_{i} \rightarrow \ell_{j} \gamma$  has the form~\cite{Borzumati:1986qx,Gabbiani:1996hi,Hisano:1995nq,Hisano:1995cp,Hisano:2007cz,Hisano:2009ae,Altmannshofer:2009ne}
\begin{equation}
T=m_{\ell_{i}} \epsilon^{\lambda} \bar{u}_{j}(p-q)\left[i q^{\nu} \sigma_{\lambda \nu}\left(A_{L} P_{L}+A_{R} P_{R}\right)\right] u_{i}(p)\,,
\end{equation}
where $p$ and $q$ are momenta of the leptons $\ell_i$ and photon respectively, $P_{R,L} = \frac{1}{2}(1\pm \gamma_5)$ and
$A_{L,R}$ are the two possible amplitudes entering the process. The lepton mass factor $m_{\ell_i}$ is associated
to the chirality flip present in this transition.  The branching ratio of $\ell_{i} \rightarrow \ell_{j} \gamma$ can be written as
\begin{equation}
\frac{\mathrm{BR}\left(\ell_{i} \rightarrow \ell_{j} \gamma\right)}{\mathrm{BR}\left(\ell_{i} \rightarrow \ell_{j} \nu_{i} \overline{\nu_{j}}\right)}=\frac{48 \pi^{3} \alpha_e}{G_{F}^{2}}\left(\left|A_{L}^{i j}\right|^{2}+\left|A_{R}^{i j}\right|^{2}\right)\,,
\end{equation}
where $\alpha_e$ is the elecromagnetic fine-structure constant and $G_F$ is the Fermi coupling constant. In the mass insertion approximation, the amplitudes read as~\cite{Paradisi:2005fk,Ciuchini:2007ha}

\begin{equation}
\label{eq:ALAR}
\begin{aligned}
                A_{L}^{i j}=& \frac{\alpha_{2}}{4 \pi} \frac{\left(\delta^{\ell}_{LL}\right)_{ij}}{m_{\tilde{\ell}}^{2}}\left[f_{1 n}\left(a_{2}\right)+f_{1 c}\left(a_{2}\right)+\frac{\mu M_{2} \tan \beta}{\left(M_{2}^{2}-\mu^{2}\right)}\left(f_{2 n}\left(a_{2}, b\right)+f_{2 c}\left(a_{2}, b\right)\right)\right.\\
                &\left.+\tan^{2}\theta_{W}\left(f_{1 n}\left(a_{1}\right)+\mu M_{1} \tan \beta\left(\frac{f_{3 n}\left(a_{1}\right)}{m_{\tilde{\ell}}^{2}}+\frac{f_{2 n}\left(a_{1}, b\right)}{\left(\mu^{2}-M_{1}^{2}\right)}\right)\right)\right] \\
                &+ \frac{\alpha_{1}}{4 \pi} \frac{\left(\delta^{\ell}_{RL}\right)_{ij}}{m_{\tilde{\ell}}^{2}}\left(\frac{M_{1}}{m_{\ell_{i}}}\right) 2 f_{2 n}\left(a_{1}\right), \\
                A_{R}^{i j}=&\frac{\alpha_{1}}{4 \pi} \left\{\frac{\left(\delta^{\ell}_{RR}\right)_{ij}}{m_{\tilde{\ell}}^{2}}\left[4 f_{1 n}\left(a_{1}\right)+\mu M_{1} \tan \beta\left(\frac{f_{3 n}\left(a_{1}\right)}{m_{\tilde{\ell}}^{2}}-\frac{2 f_{2 n}\left(a_{1}, b\right)}{\left(\mu^{2}-M_{1}^{2}\right)}\right)\right]\right.\\
                &\left.+\frac{\left(\delta^{\ell}_{LR}\right)_{ij}}{m_{\tilde{\ell}}^{2}}\left(\frac{M_{1}}{m_{\ell_{i}}}\right) 2 f_{2 n}\left(a_{1}\right)\right\},
        \end{aligned}
\end{equation}
where $\theta_W$ is the weak mixing angle and $m_{\ell_i}$ is the charged lepton mass, $a_{1,2} = M^2_{1,2}/m_{\tilde{\ell}}^{2}$, $b = \mu^2/m_{\tilde{\ell}}^{2} $ and $f_{i(c,n)}(x,y) = f_{i(c,n)}(x) -
f_{i(c,n)}(y)$. The parameter $\mu$ is given through the requirement of the correct electrweak symmetry breaking, at low energy scale we have~\cite{Masina:2002mv,Feruglio:2009hu},
\begin{equation}
|\mu|^2\simeq m_0^2\frac{1+0.5\tan^2\beta}{\tan^2\beta-1}+M^2_{1/2}\frac{0.5+3.5\tan^2\beta}{\tan^2\beta-1}-\frac{1}{2}m_Z^2\,.
\end{equation}
The loop functions $f_i$ are given as~\cite{Paradisi:2005fk,Ciuchini:2007ha}
\begin{equation}
 \begin{aligned}
&f_{1 n}(x)=\left(-17 x^{3}+9 x^{2}+9 x-1+6 x^{2}(x+3) \ln x\right) /\left(24(1-x)^{5}\right)\,, \\
&f_{2 n}(x)=\left(-5 x^{2}+4 x+1+2 x(x+2) \ln x\right) /\left(4(1-x)^{4}\right)\,, \\
&f_{3 n}(x)=\left(1+9 x-9 x^{2}-x^{3}+6 x(x+1) \ln x\right) \textit{}/\left(3(1-x)^{5}\right)\,, \\
&f_{1 c}(x)=\left(-x^{3}-9 x^{2}+9 x+1+6 x(x+1) \ln x\right) /\left(6(1-x)^{5}\right) \,,\\
&f_{2 c}(x)=\left(-x^{2}-4 x+5+2(2 x+1) \ln x\right) /\left(2(1-x)^{4}\right)\,.
 \end{aligned}
\end{equation}
As we mentioned above, $\delta^{\ell}_{LL}$ and $\delta^{\ell}_{RR}$ still have no off-diagonal terms in the SCKM basis, so the contribution to $\mu\to e\gamma$ branching ratio come only from the terms of $\delta^{\ell}_{LR}$ and $\delta^{\ell}_{RL}$ in Eq.~\eqref{eq:ALAR}. In numerical calculations of the $\mu\to e\gamma$ branching ratio, the input parameters contain $m_{3/2},m_0,F^\tau,M_{1/2}$, while the flavor parameters in the slepton mass matrices have been fixed to the best-fit values, i.e. Eq.~\eqref{eq:SleptonMass}, and $\tan\beta=5$. We expect that the SUSY breaking parameter $F^\tau$ to be the same order as $m_0$ and $m_{3/2}$, and in order to prevent the tachyonic slepton, we take $F^\tau=m_0/4\approx m_{3/2}/4$.  After fixing the value of $M_{1/2}$, the $\mu\to e\gamma$ ratio only depends on the slepton mass scale $m_0$, we plot $\mathrm{BR}(\mu\to e\gamma)$ versus  $m_0$ in figure~\ref{fig:LFV} for $M_{1/2}=5,10,15~\mathrm{TeV}$.
As you can see, the predicted $\mathrm{BR}(\mu\to e\gamma)$ is lower than the experimental upper bound as far as the gaugino mass scale $M_{1/2}$ is larger than $10~\mathrm{TeV}$, while when $M_{1/2}=5~\mathrm{TeV}$ the SUSY mass scale $m_0$ should be large than around $5~\mathrm{TeV}$ to be consistent with the current bound $\mathrm{BR}(\mu\to e\gamma)< 4.2\times 10^{-13}$~\cite{MEG:2016leq}.

\begin{figure}[ht!]
\centering
\includegraphics[width=0.85\textwidth]{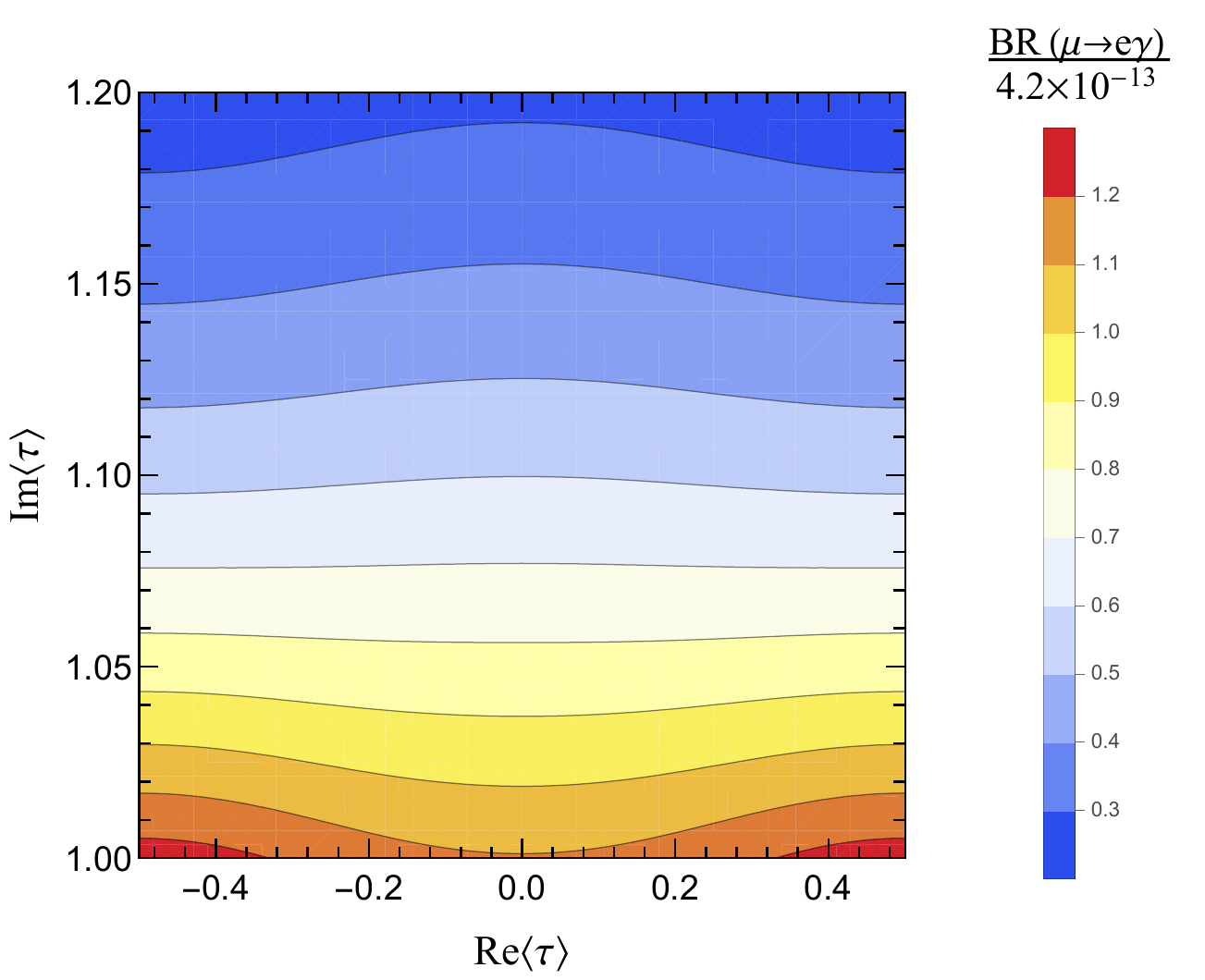}
\caption{The contour plot of $\mathrm{BR}(\mu\to e\gamma)$ in the $\tau$ plane for $m_0=M_{1/2}=5~\mathrm{TeV},F^\tau=m_0/4$. The value of branching ratio is normalized with the current upper bound $4.2\times 10^{-13}$~\cite{MEG:2016leq}. The other free parameters are fixed at their best-fit values as given in Eq.~\eqref{eq:best-fit values}. }
\label{fig:Contour}
\end{figure}

On the other hand, if we fix the SUSY parameters $m_0, M_{1/2}$ and flavor parameters $\alpha,\beta,\gamma, g^2/\Lambda$, while let $\langle\tau\rangle$ be freely distributed in fundamental
domain, so that we can obtain a contour map of $\mathrm{BR}(\mu\to e\gamma)$ in the $\tau$ plane, as shown in figure~\ref{fig:Contour}.
It is clear that $\mathrm{BR}(\mu\to e\gamma)$  is more sensitive to $\mathrm{Im}\langle \tau\rangle$ and less sensitive to $\mathrm{Re}\langle \tau\rangle$, in particular, $\mathrm{BR}(\mu\to e\gamma)$ decreased significantly with the increase of $\mathrm{Im}\langle \tau\rangle$. Moreover, $\mathrm{BR}(\mu\to e\gamma)$ is below the current bound when $\mathrm{Im}\langle \tau\rangle>1.05$.

Finally, we can also discuss other LFV processes such as $\ell_i\rightarrow\ell_{j}\ell_{k}\bar{\ell}_k$ and $\mu N\rightarrow eN$ in nuclei. These channels are typically dominated by the dipole operators in SUSY models, leading to the following simple relations~\cite{Altmannshofer:2009ne}:
\begin{equation}
\begin{aligned}
&\mathrm{BR}\left(\ell_{i} \rightarrow \ell_{j} \ell_{k} \bar{\ell}_{k}\right) \simeq \frac{\alpha_{e}}{3 \pi}\left(\log \frac{m_{\ell_{i}}^{2}}{m_{\ell_{k}}^{2}}-3\right) \mathrm{BR}\left(\ell_{i} \rightarrow \ell_{j} \gamma\right)\,,\\
&\mathrm{CR}(\mu N\rightarrow  e N) \simeq \alpha_{e} \mathrm{BR}(\mu \rightarrow e \gamma)\,.
\end{aligned}
\end{equation}
The numerical results can be directly obtained, so we omit the detailed discussion in this part. It is only necessary to mention that for $m_0=4F^\tau=M_{1/2}=5\,\text{TeV}$, both branching ration $\mathrm{BR}\left(\mu \rightarrow e e \bar{e}\right)$ and conversion rate $\mathrm{CR}(\mu N\rightarrow  e N)$ are roughly $\mathcal{O}(10^{-15})$, which are clearly below their respective current experimental bounds $\mathrm{BR}\left(\mu \rightarrow e e \bar{e}\right)< 1.0\times 10^{-12}$ and $\mathrm{CR}(\mu N\rightarrow  e N)<7.0\times 10^{-13}$~\cite{ParticleDataGroup:2020ssz}.

\section{\label{sec:summary}Summary and conclusions}

In this paper, we find a modular neutrino model with the fewest input parameters so far. This model is based on the $S'_4$ modular symmetry in combination with gCP symmetry. It uses four coupling constants $\beta$, $\gamma$, $g^2/\Lambda$ and the complex modulus $\tau$ to well explain the  three charged lepton masses, the three light neutrino masses, the three neutrino mixing angles and the three CP violation phases. From the numerical scan of the parameter space, we find that all the mixing angles and CP violation phases vary in very small regions. In particular, the atmospheric mixing angel and the Dirac CP phase is predicted lie in the range $\sin^2\theta_{23}\in[0.504, 0.510]$ and $\delta_{CP}\in[1.316\pi, 1.364\pi]$ respectively. All the predictions of our model are compatible with the experimental data from neutrino oscillation, tritium decay, neutrinoless double decay and cosmology. We expect the models could be tested at future neutrino facilities and ton scale neutrinoless double beta decay experiments.

We also discuss the LFV phenomenology of this model in the moduli-mediated SUSY breaking framework, where the soft SUSY breaking terms arise from the modulus $F-$term in the modular neutrino model.  These soft breaking terms also have a certain flavor structure because they are constrained to be non-holomorphic modular forms due to the modular symmetry. We have studied the dependence of the branching ratio $\mathrm{BR}(\mu\to e\gamma)$ on the slepton mass scale $m_0$, gauginos mass scale $M_{1/2}$ and modulus VEV $\langle\tau\rangle$, and we find that the $\mathrm{BR}(\mu\to e\gamma)$ is always below the current bound when the gaugino mass scale $M_{1/2}$ is larger than $10~\mathrm{TeV}$. On the other hand, the branching ratio $\mathrm{BR}(\mu\to e\gamma)$ also depends significantly on the moduli vacuum as shown in figure~\ref{fig:Contour}, where the $\mathrm{BR}(\mu\to e\gamma)$ decreased rapidly with the increase of $\mathrm{Im}\langle \tau\rangle$. A similar analysis can be fully implemented in other lepton flavor violation processes, such as $\tau\to e\gamma$ and $\tau\to \mu \gamma$.

\section*{Acknowledgements}

GJD and XGL are supported by the National Natural Science Foundation of China under Grant Nos.~11975224, 11835013. XGL is also supported in part by U.S. National Science Foundation under Grant No.~PHY-2210283. CYY is supported in part by the Grants No.~NSFC-11975130, No.~NSFC-12035008, No.~NSFC-12047533, the Helmholtz-OCPC International Postdoctoral Exchange Fellowship Program, the National Key Research and Development Program of China under Grant No.~2017YFA0402200, the China Postdoctoral Science Foundation under Grant No.~2018M641621, and the Deutsche Forschungsgemeinschaft (DFG, German Research Foundation) under Germany’s Excellence Strategy — EXC 2121 ``Quantum Universe'' — 390833306.

\section*{Appendix}

\setcounter{equation}{0}
\renewcommand{\theequation}{\thesection.\arabic{equation}}

\begin{appendix}

\section{\label{sec:app-S4DC}Group theory of $\Gamma'_4\cong S'_4$}
\cleqn

The homogeneous finite modular group $\Gamma'_4\cong S'_4$ has 48 elements, and it can be generated by three generators $S,T$ and $R$ obeying the rules:
\begin{equation}
S^2=R,\quad (ST)^3=T^4=R^2=1,\quad TR=RT\,.
\end{equation}
Its group \texttt{ID} in \texttt{GAP}~\cite{GAP} is [48, 30]. Notice that $S_4$ is not a subgroup of $S'_4$, it is isomorphic to the quotient group of $S'_4$ over $Z_2^{R}$, i.e. $S_4\cong S'_4/Z_2^{R}$, where $Z_2^{R}=\{1,R\}$ is the center and a normal subgroup of $S'_4$. The finite modular group $S'_4$ is a  double cover of $S_4$.
It is notable that $S'_4$ is  isomorphic to the semidirect product of $A_4$ with $Z_4$, namely $S'_4 \cong A_4 \rtimes Z_4$.
In other words, $S'_4$ can also be regarded as a split extension of $A_4$ by $Z_4$.

The group $S'_4$ has four singlet representations $\mathbf{1},\mathbf{1}^{\prime},\mathbf{\widehat{1}}$ and $\mathbf{\hat{1}}^{\prime}$, two doublet representations $\mathbf{2}$ and $\mathbf{\widehat{2}}$, and four triplet representations $\mathbf{3},\mathbf{3}^{\prime},\mathbf{\widehat{3}}$ and $\mathbf{\widehat{3}}^{\prime}$. We present the representation matrices of the generators in different irreducible representations in table~\ref{tab:Rep_matrix}. In the representations $\mathbf{1}$, $\mathbf{1}^{\prime}$, $\mathbf{2}$, $\mathbf{3}$ and $\mathbf{3}^{\prime}$, the generator $R=1$ is identity matrix, the representation matrices of $S$ and $T$ coincide with those of $S_4$~\cite{Gui-JunDing:2019wap}, consequently $S'_4$ can not be distinguished from $S_4$ in these representations. In the hatted representations $\mathbf{\hat{1}}$, $\mathbf{\widehat{1}}^{\prime}$, $\mathbf{\widehat{2}}$, $\mathbf{\widehat{3}}$ and $\mathbf{\widehat{3}}^{\prime}$, we have the generator $R=-1$.

\begin{table}[!t]
\centering
\begin{tabular}{|c|c|c|c|}\hline\hline
 & $S$ & $T$ & $R$\\ \hline
  $\mathbf{1},\mathbf{1^{\prime}}$ & $\pm 1$ & $\pm 1$ & $1$ \\ \hline
  $\mathbf{\widehat{1}},\mathbf{\widehat{1}^{\prime}}$ & $\pm i$ & $\mp i$ & $-1$ \\ \hline
  $\mathbf{2}$ & $\dfrac{1}{2}\begin{pmatrix}
  -1 & \sqrt{3} \\
 \sqrt{3} & 1 \\
\end{pmatrix}$ & $\begin{pmatrix}
 1 & 0 \\
 0 & -1 \\
\end{pmatrix}$ & $\begin{pmatrix}{1} & {0} \\ {0} & {1}\end{pmatrix}$ \\ \hline
  $\mathbf{\widehat{2}}$ & $\dfrac{i}{2}\begin{pmatrix}
 -1 & \sqrt{3} \\
 \sqrt{3} & 1 \\
\end{pmatrix}$ & $-i\begin{pmatrix}
 1 & 0 \\
 0 & -1 \\
\end{pmatrix}$ & $-\begin{pmatrix}{1} & {0} \\ {0} & {1}\end{pmatrix}$ \\ \hline
    $\mathbf{3},\mathbf{3^{\prime}}$ & $\pm\dfrac{1}{2}\begin{pmatrix}
 0 & \sqrt{2} & \sqrt{2} \\
 \sqrt{2} & -1 & 1 \\
 \sqrt{2} & 1 & -1 \\
\end{pmatrix}$ & $\pm\begin{pmatrix}
 1 & 0 & 0 \\
 0 & i & 0 \\
 0 & 0 & -i \\
\end{pmatrix}$ & $\begin{pmatrix}{1} & {0} & {0} \\ {0} & {1} & {0} \\ {0} & {0} & {1}\end{pmatrix}$ \\ \hline
  $\mathbf{\widehat{3}}, \mathbf{\widehat{3}^{\prime}}$ & $\pm\dfrac{i}{2}\begin{pmatrix}
 0 & \sqrt{2} & \sqrt{2} \\
 \sqrt{2} & -1 & 1 \\
 \sqrt{2} & 1 & -1 \\
\end{pmatrix}$ & $\mp i\begin{pmatrix}
 1 & 0 & 0 \\
 0 & i & 0 \\
 0 & 0 & -i \\
\end{pmatrix}$ & $-\begin{pmatrix}{1} & {0} & {0} \\ {0} & {1} & {0} \\ {0} & {0} & {1}\end{pmatrix}$ \\ \hline\hline
\end{tabular}
\caption{The representation matrices of the generators $S, T$ and $R$ for different irreducible representations of $S'_{4}$ in the $T$-diagonal basis. }
\label{tab:Rep_matrix}
\end{table}
The tensor products between irreducible representations and the Clebsch-Gordan coefficients of $S'_4$ are needed when constructing a concrete $S'_4$ model. In the following, all CG coefficients are given in the form of $\alpha\otimes\beta$, we use $\alpha_i(\beta_i)$ to denote the component of the left (right) basis vector $\alpha (\beta)$. The notations $\mathbf{I}$, $\mathbf{II}$,  $\mathbf{III}$ and $\mathbf{IV}$ stand for singlet, doublet, triplet and quartet representations of $S'_4$ respectively.

\begin{itemize}
\item $\mathbf{I}\otimes\mathbf{I}\to\mathbf{I}$\,,
\begin{equation}
\begin{array}{ll}
  \left.\begin{array}{l}
\mathbf{1}\otimes\mathbf{1}\to\mathbf{1_{s}},~~
\mathbf{1}\otimes\mathbf{1'}\to\mathbf{1'}\\
\mathbf{1}\otimes\mathbf{\widehat{1}}\to\mathbf{\widehat{1}},~~~
\mathbf{1}\otimes\mathbf{\widehat{1}'}\to\mathbf{\widehat{1}'}\\
\mathbf{1'}\otimes\mathbf{1'}\to\mathbf{1_{s}},~
\mathbf{1'}\otimes\mathbf{\widehat{1}}\to\mathbf{\widehat{1}'}\\
\mathbf{1'}\otimes\mathbf{\widehat{1}'}\to\mathbf{\widehat{1}},~~
\mathbf{\widehat{1}}\otimes\mathbf{\widehat{1}}\to\mathbf{1'_{s}}\\
\mathbf{\widehat{1}}\otimes\mathbf{\widehat{1}'}\to\mathbf{1},~~~
\mathbf{\widehat{1}'}\otimes\mathbf{\widehat{1}'}\to\mathbf{1'_{s}}
  \end{array}\right\} &~~~
\begin{array}{l}
\mathbf{I} \sim \alpha\beta
\end{array}
\end{array}\nonumber
\end{equation}
\end{itemize}

\begin{itemize}
\item $\mathbf{I}\otimes\mathbf{II}\to\mathbf{II}$\,,
\begin{equation}
\begin{array}{lll}
\begin{array}{c}
    ~  \\ [-1ex]n=0  \\ \\ \\[1ex]n=1 \\ \\
  \end{array} ~~~~&
  \left.\begin{array}{l}
 \mathbf{1}\otimes\mathbf{2}\to\mathbf{2},~~~
\mathbf{1}\otimes\mathbf{\widehat{2}}\to\mathbf{\widehat{2}}\\
\mathbf{\widehat{1}}\otimes\mathbf{2}\to\mathbf{\widehat{2}},~~~
\mathbf{\widehat{1}'}\otimes\mathbf{\widehat{2}}\to\mathbf{2}\\ \\
\mathbf{1'}\otimes\mathbf{2}\to\mathbf{2},~~~
\mathbf{1'}\otimes\mathbf{\widehat{2}}\to\mathbf{\widehat{2}}\\
\mathbf{\widehat{1}}\otimes\mathbf{\widehat{2}}\to\mathbf{2},~~~
\mathbf{\widehat{1}'}\otimes\mathbf{2}\to\mathbf{\widehat{2}}
  \end{array}\right\} &~~~
\begin{array}{l}
\mathbf{II} \sim  \alpha M^{(n)}\begin{pmatrix}
\beta_1\\
\beta_2
\end{pmatrix}\\
\end{array}
\end{array}\nonumber
\end{equation}
\end{itemize}

where $M^{(0)}=
\begin{pmatrix}
  1 ~&~ 0\\0 ~&~ 1
\end{pmatrix}
$, $M^{(1)}=
\begin{pmatrix}
  0 ~&~ 1\\-1 ~&~ 0
\end{pmatrix}
$, it's the same below.

\begin{itemize}
\item $\mathbf{I}\otimes\mathbf{III}\to\mathbf{III}$\,,
\begin{equation}
\begin{array}{ll}
 \left.\begin{array}{l}
\mathbf{1}\otimes\mathbf{3}\to\mathbf{3},~~~
\mathbf{1}\otimes\mathbf{3'}\to\mathbf{3'}\\
\mathbf{1}\otimes\mathbf{\widehat{3}}\to\mathbf{\widehat{3}},~~~
\mathbf{1}\otimes\mathbf{\widehat{3}'}\to\mathbf{\widehat{3}'}\\
\mathbf{1'}\otimes\mathbf{3}\to\mathbf{3'},~~
\mathbf{1'}\otimes\mathbf{3'}\to\mathbf{3}\\
\mathbf{1'}\otimes\mathbf{\widehat{3}}\to\mathbf{\widehat{3}'},~~
\mathbf{1'}\otimes\mathbf{\widehat{3}'}\to\mathbf{\widehat{3}}\\
\mathbf{\widehat{1}}\otimes\mathbf{3}\to\mathbf{\widehat{3}},~~~\,
\mathbf{\widehat{1}}\otimes\mathbf{3'}\to\mathbf{\widehat{3}'}\\
\mathbf{\widehat{1}}\otimes\mathbf{\widehat{3}}\to\mathbf{3'},~~~
\mathbf{\widehat{1}}\otimes\mathbf{\widehat{3}'}\to\mathbf{3}\\
\mathbf{\widehat{1}'}\otimes\mathbf{3}\to\mathbf{\widehat{3}'},~~\,
\mathbf{\widehat{1}'}\otimes\mathbf{3'}\to\mathbf{\widehat{3}}\\
\mathbf{\widehat{1}'}\otimes\mathbf{\widehat{3}}\to\mathbf{3},~~~
\mathbf{\widehat{1}'}\otimes\mathbf{\widehat{3}'}\to\mathbf{3'}
  \end{array}\right\} &~~~
\begin{array}{l}
\mathbf{III} \sim \alpha\begin{pmatrix}
\beta_1\\
\beta_2\\
\beta_3
\end{pmatrix}
\end{array}
\end{array}\nonumber
\end{equation}
\end{itemize}

\begin{itemize}
\item $\mathbf{II}\otimes\mathbf{II}\to\mathbf{I}_1\oplus\mathbf{I}_2\oplus\mathbf{II}$\,,
\begin{equation}
\begin{array}{lll}
\begin{array}{c}
    ~\\ [-1.1ex]n=0  \\ \\[1ex]n=1 \\
  \end{array} ~~~~&
  \left.\begin{array}{l}
\mathbf{2}\otimes\mathbf{2}\to\mathbf{1'_{a}}\oplus\mathbf{1_{s}}\oplus\mathbf{2_{s}}\\
\mathbf{2}\otimes\mathbf{\widehat{2}}\to\mathbf{\widehat{1}'}\oplus\mathbf{\widehat{1}}\oplus\mathbf{\widehat{2}}\\\\
\mathbf{\widehat{2}}\otimes\mathbf{\widehat{2}}\to\mathbf{1_{a}}\oplus\mathbf{1'_{s}}\oplus\mathbf{2_{s}}
  \end{array}\right\} &~~~
\begin{array}{l}
\mathbf{I}_1 \sim \alpha_1\beta_2-\alpha_2\beta_1\\[1ex]
\mathbf{I}_2 \sim \alpha_1\beta_1+\alpha_2\beta_2\\[1ex]
\mathbf{II} \sim M^{(n)}\begin{pmatrix}
-\alpha_1\beta_1+\alpha_2\beta_2\\
\alpha_1\beta_2+\alpha_2\beta_1
\end{pmatrix}
\end{array}
\end{array}\nonumber
\end{equation}
\end{itemize}

\begin{itemize}
\item $\mathbf{II}\otimes\mathbf{III}\to\mathbf{III}_1\oplus\mathbf{III}_2$\,,
\begin{equation}
\begin{array}{ll}
 \left.\begin{array}{l}
\mathbf{2}\otimes\mathbf{3}\to\mathbf{3}\oplus\mathbf{3'}\\
\mathbf{2}\otimes\mathbf{3'}\to\mathbf{3'}\oplus\mathbf{3}\\
\mathbf{2}\otimes\mathbf{\widehat{3}}\to\mathbf{\widehat{3}}\oplus\mathbf{\widehat{3}'}\\
\mathbf{2}\otimes\mathbf{\widehat{3}'}\to\mathbf{\widehat{3}'}\oplus\mathbf{\widehat{3}}\\
\mathbf{\widehat{2}}\otimes\mathbf{3}\to\mathbf{\widehat{3}}\oplus\mathbf{\widehat{3}'}\\
\mathbf{\widehat{2}}\otimes\mathbf{3'}\to\mathbf{\widehat{3}'}\oplus\mathbf{\widehat{3}}\\
\mathbf{\widehat{2}}\otimes\mathbf{\widehat{3}}\to\mathbf{3'}\oplus\mathbf{3}\\
\mathbf{\widehat{2}}\otimes\mathbf{\widehat{3}'}\to\mathbf{3}\oplus\mathbf{3'}
  \end{array}\right\} &~~~
\begin{array}{l}
\mathbf{III}_1 \sim \begin{pmatrix}
2\alpha_1\beta_1\\
-\alpha_1\beta_2+\sqrt{3}\alpha_2\beta_3\\
-\alpha_1\beta_3+\sqrt{3}\alpha_2\beta_2
\end{pmatrix}\\[5ex]
\mathbf{III}_2 \sim \begin{pmatrix}
-2\alpha_2\beta_1\\
\sqrt{3}\alpha_1\beta_3+\alpha_2\beta_2\\
\sqrt{3}\alpha_1\beta_2+\alpha_2\beta_3
\end{pmatrix}
\end{array}
\end{array}\nonumber
\end{equation}
\end{itemize}

\begin{itemize}
\item $\mathbf{III}\otimes\mathbf{III}\to\mathbf{I}\oplus\mathbf{II}\oplus\mathbf{III}_1\oplus\mathbf{III}_2$\,,
\begin{equation}
\begin{array}{lll}
\begin{array}{c}
    ~\\ [-3ex]n=0  \\ \\ \\ \\[3ex]n=1 \\
  \end{array} ~~~~&
  \left.\begin{array}{l}
\mathbf{3}\otimes\mathbf{3}\to\mathbf{1_{s}}\oplus\mathbf{2_{s}}\oplus\mathbf{3_{a}}\oplus\mathbf{3'_{s}}\\
\mathbf{3}\otimes\mathbf{\widehat{3}}\to\mathbf{\widehat{1}}\oplus\mathbf{\widehat{2}}\oplus\mathbf{\widehat{3}}\oplus\mathbf{\widehat{3}'}\\
\mathbf{3'}\otimes\mathbf{3'}\to\mathbf{1_{s}}\oplus\mathbf{2_{s}}\oplus\mathbf{3_{a}}\oplus\mathbf{3'_{s}}\\
\mathbf{3'}\otimes\mathbf{\widehat{3}'}\to\mathbf{\widehat{1}}\oplus\mathbf{\widehat{2}}\oplus\mathbf{\widehat{3}}\oplus\mathbf{\widehat{3}'}\\
\mathbf{\widehat{3}}\otimes\mathbf{\widehat{3}'}\to\mathbf{1}\oplus\mathbf{2}\oplus\mathbf{3}\oplus\mathbf{3'}\\\\
\mathbf{3}\otimes\mathbf{3'}\to\mathbf{1'}\oplus\mathbf{2}\oplus\mathbf{3'}\oplus\mathbf{3}\\
\mathbf{3}\otimes\mathbf{\widehat{3}'}\to\mathbf{\widehat{1}'}\oplus\mathbf{\widehat{2}}\oplus\mathbf{\widehat{3}'}\oplus\mathbf{\widehat{3}}\\
\mathbf{3'}\otimes\mathbf{\widehat{3}}\to\mathbf{\widehat{1}'}\oplus\mathbf{\widehat{2}}\oplus\mathbf{\widehat{3}'}\oplus\mathbf{\widehat{3}}\\
\mathbf{\widehat{3}}\otimes\mathbf{\widehat{3}}\to\mathbf{1'_{s}}\oplus\mathbf{2_{s}}\oplus\mathbf{3'_{a}}\oplus\mathbf{3_{s}}\\
\mathbf{\widehat{3}'}\otimes\mathbf{\widehat{3}'}\to\mathbf{1'_{s}}\oplus\mathbf{2_{s}}\oplus\mathbf{3'_{a}}\oplus\mathbf{3_{s}}
  \end{array}\right\} &~~~
\begin{array}{l}
\mathbf{I} \sim \alpha_1\beta_1+\alpha_2\beta_3+\alpha_3\beta_2\\[3ex]
\mathbf{II} \sim M^{(n)}\begin{pmatrix}
2\alpha_1\beta_1-\alpha_2\beta_3-\alpha_3\beta_2\\
\sqrt{3}\alpha_2\beta_2+\sqrt{3}\alpha_3\beta_3
\end{pmatrix}\\[3ex]
\mathbf{III}_1 \sim \begin{pmatrix}
\alpha_2\beta_3-\alpha_3\beta_2\\
\alpha_1\beta_2-\alpha_2\beta_1\\
-\alpha_1\beta_3+\alpha_3\beta_1
\end{pmatrix}\\[5ex]
\mathbf{III}_2 \sim \begin{pmatrix}
\alpha_2\beta_2-\alpha_3\beta_3\\
-\alpha_1\beta_3-\alpha_3\beta_1\\
\alpha_1\beta_2+\alpha_2\beta_1
\end{pmatrix}
\end{array}
\end{array}\nonumber
\end{equation}
\end{itemize}

\section{\label{sec:MF}Integer weight modular forms of level 4}

The structure of the modular forms space of weight $k$ and level 4 (non-negative integer or half-integer) is well known, and it can be constructed by making use of the theta constants ~\cite{Liu:2020msy}:
\begin{align}
\label{eq:Mk_Gamma4}
\nonumber
\mathcal{M}_{k}(\Gamma(4))&=\bigoplus_{a+b=2k,\, a,b\ge 0} \mathbb{C} \theta^{a}_2(\tau)\theta^{b}_3(\tau)\,,
\end{align}
where the theta constants is defined as
\begin{eqnarray}
\nonumber \theta_2(\tau)&=& \sum_{m\in\mathbb{Z}} e^{2\pi i \tau (m+1/2)^2}=2q^{1/4}(1+q^2+q^6+q^{12}+\dots)\,,\\
\theta_3(\tau)&=&\sum_{m\in\mathbb{Z}} e^{2\pi i \tau m^2}=1+2q+2q^4+2q^9+2q^{16}+\dots \,.
\end{eqnarray}
The weight $k$ modular forms of level 4 can be expressed as the homogeneous polynomials of degree $2k$ in $\theta_{1}$ and $\theta_{2}$. Consequently the linear space of weight $k$ and level 4 modular forms has dimension $2k+1$. In the following, we report the explicit expressions of the $S'_4$ modular multiplets up to weight 6 in our working basis summarized in table~\ref{tab:Rep_matrix}. We prefer to use $\vartheta_1(\tau)=\theta_3(\tau)$, $\vartheta_2(\tau)=-\theta_2(\tau)$ since $\vartheta_1(\tau)$ and $\vartheta_2(\tau)$ turns out to be half weight modular forms and they form a doublet of the metaplectic cover of $S'_4$~\cite{Liu:2020msy}.
\begin{itemize}
\item{$k_Y=1$}
\begin{equation}
        Y_{\mathbf{\widehat{3}'}}^{(1)}=\begin{pmatrix}
                \sqrt{2}\vartheta_1\vartheta_2\\
                -\vartheta_2^2\\
                \vartheta_1^2\\
        \end{pmatrix}\,.
\end{equation}

\item{$k_Y=2$}

\begin{eqnarray}
        \nonumber&&Y_{\mathbf{2}}^{(2)}=\begin{pmatrix}
                \vartheta_1^4+\vartheta_2^4\\
                -2\sqrt{3}\vartheta_1^2\vartheta_2^2\\
        \end{pmatrix}\,,\\
        &&Y_{\mathbf{3}}^{(2)}=\begin{pmatrix}
                \vartheta_1^4-\vartheta_2^4\\
                2\sqrt{2}\vartheta_1^3\vartheta_2\\
                2\sqrt{2}\vartheta_1\vartheta_2^3\\
        \end{pmatrix}\,.
\end{eqnarray}

\item{$k_Y=3$}

\begin{eqnarray}
        \nonumber&&Y_{\mathbf{\widehat{1}'}}^{(3)}=\vartheta_1\vartheta_2\left(\vartheta_1^4-\vartheta_2^4\right)\,, \\
        \nonumber&&Y_{\mathbf{\widehat{3}}}^{(3)}=\begin{pmatrix}
                4\sqrt{2}\vartheta_1^3\vartheta_2^3\\
                \vartheta_1^6+3\vartheta_1^2\vartheta_2^4\\
                -\vartheta_2^2\left(3\vartheta_1^4+\vartheta_2^4\right)\\
        \end{pmatrix}\,,\\
        &&Y_{\mathbf{\widehat{3}'}}^{(3)}=\begin{pmatrix}
                2\sqrt{2}\vartheta_1\vartheta_2\left(\vartheta_1^4+\vartheta_2^4\right)\\
                \vartheta_2^6-5\vartheta_1^4\vartheta_2^2\\
                5\vartheta_1^2\vartheta_2^4-\vartheta_1^6\\
        \end{pmatrix}\,.
\end{eqnarray}

\item{$k_Y=4$}

\begin{eqnarray}
        \nonumber&&Y_{\mathbf{1}}^{(4)}=
        \vartheta_1^8+14\vartheta_1^4\vartheta_2^4+\vartheta_2^8\,,\\
        \nonumber&&Y_{\mathbf{2}}^{(4)}=\begin{pmatrix}
                \vartheta_1^8-10\vartheta_1^4\vartheta_2^4+\vartheta_2^8\\
                4\sqrt{3}\vartheta_1^2\vartheta_2^2\left(\vartheta_1^4+\vartheta_2^4\right)\\
        \end{pmatrix}\,,\\
        \nonumber&&Y_{\mathbf{3}}^{(4)}=\begin{pmatrix}
                \vartheta_2^8-\vartheta_1^8\\
                \sqrt{2}\vartheta_2\left(\vartheta_1^7+7\vartheta_1^3\vartheta_2^4\right)\\
                \sqrt{2}\vartheta_1\left(\vartheta_2^7+7\vartheta_1^4\vartheta_2^3\right)\\
        \end{pmatrix}\,,\\
        &&Y_{\mathbf{3'}}^{(4)}=\vartheta_1\vartheta_2\left(\vartheta_1^4-\vartheta_2^4\right)\begin{pmatrix}
                \sqrt{2}\vartheta_1\vartheta_2\\
                -\vartheta_2^2\\
                \vartheta_1^2\\
        \end{pmatrix}\,.
\end{eqnarray}

\item{$k_Y=5$}

\begin{eqnarray}
        \nonumber&&Y_{\mathbf{\widehat{2}}}^{(5)}=\vartheta_1\vartheta_2\left(\vartheta_1^4-\vartheta_2^4\right)\begin{pmatrix}
                2\sqrt{3}\vartheta_1^2\vartheta_2^2\\
                \vartheta_1^4+\vartheta_2^4\\
        \end{pmatrix}\,, \\
        \nonumber&&Y_{\mathbf{\widehat{3}}}^{(5)}=\begin{pmatrix}
                -8\sqrt{2}\vartheta_1^3\vartheta_2^3\left(\vartheta_1^4+\vartheta_2^4\right)\\
                \vartheta_1^2\left(\vartheta_1^8-14\vartheta_1^4\vartheta_2^4-3\vartheta_2^8\right)\\
                \vartheta_2^2\left(3\vartheta_1^8+14\vartheta_1^4\vartheta_2^4-\vartheta_2^8\right)\\
        \end{pmatrix}\,, \\
        \nonumber&&Y_{\mathbf{\widehat{3}'}I}^{(5)}=\begin{pmatrix}
                2\sqrt{2}\vartheta_1\vartheta_2\left(\vartheta_1^8-10\vartheta_1^4\vartheta_2^4+\vartheta_2^8\right)\\
                \vartheta_2^2\left(13\vartheta_1^8+2\vartheta_1^4\vartheta_2^4+\vartheta_2^8\right)\\
                -\vartheta_1^2\left(\vartheta_1^8+2\vartheta_1^4\vartheta_2^4+13\vartheta_2^8\right)\\
        \end{pmatrix}\,, \\
        &&Y_{\mathbf{\widehat{3}'}II}^{(5)}=\left(\vartheta_1^8+14\vartheta_1^4\vartheta_2^4+\vartheta_2^8\right)\begin{pmatrix}
                \sqrt{2}\vartheta_1\vartheta_2\\
                -\vartheta_2^2\\
                \vartheta_1^2\\
        \end{pmatrix}\,.
\end{eqnarray}

\item{$k_Y=6$}

\begin{eqnarray}
        \nonumber&&Y_{\mathbf{1}}^{(6)}=
        \vartheta_1^{12}-33\vartheta_1^8\vartheta_2^4-33\vartheta_1^4\vartheta_2^8+\vartheta_2^{12}\,,\\
        \nonumber&&Y_{\mathbf{1'}}^{(6)}=
        \vartheta_1^2\vartheta_2^2\left(\vartheta_1^4-\vartheta_2^4\right)^2 \,, \\
        \nonumber&&Y_{\mathbf{2}}^{(6)}=\left(\vartheta_1^8+14\vartheta_1^4\vartheta_2^4+\vartheta_2^8\right)\begin{pmatrix}
                \vartheta_1^4+\vartheta_2^4\\
                -2\sqrt{3}\vartheta_1^2\vartheta_2^2\\
        \end{pmatrix}\,,\\
        \nonumber&&Y_{\mathbf{3}I}^{(6)}=\begin{pmatrix}
                \vartheta_1^{12}-11\vartheta_1^8\vartheta_2^4+11\vartheta_1^4\vartheta_2^8-\vartheta_2^{12}\\
                -\sqrt{2}\vartheta_1^3\vartheta_2\left(\vartheta_1^8-22\vartheta_1^4\vartheta_2^4-11\vartheta_2^8\right)\\
                \sqrt{2}\vartheta_1\vartheta_2^3\left(11\vartheta_1^8+22\vartheta_1^4\vartheta_2^4-\vartheta_2^8\right)\\
        \end{pmatrix}\,,\\
        \nonumber&&Y_{\mathbf{3}II}^{(6)}=\left(\vartheta_1^8+14\vartheta_2^4\vartheta_1^4+\vartheta_2^8\right)\begin{pmatrix}
                \vartheta_1^4-\vartheta_2^4\\
                2\sqrt{2}\vartheta_1^3\vartheta_2\\
                2\sqrt{2}\vartheta_1\vartheta_2^3\\
        \end{pmatrix}\,,\\
        &&Y_{\mathbf{3'}}^{(6)}=\vartheta_1\vartheta_2\left(\vartheta_1^4-\vartheta_2^4\right)\begin{pmatrix}
                2\sqrt{2}\vartheta_1\vartheta_2\left(\vartheta_1^4+\vartheta_2^4\right)\\
                \vartheta_2^6-5\vartheta_1^4\vartheta_2^2\\
                5\vartheta_1^2\vartheta_2^4-\vartheta_1^6\\
        \end{pmatrix}\,.
\end{eqnarray}
The higher weight modular forms can be constructed from the tensor products of the above modular multiplets.

\end{itemize}

\end{appendix}

\bibliographystyle{utphys}

\providecommand{\href}[2]{#2}\begingroup\raggedright\endgroup

\end{document}